\documentclass[11pt,prd,preprintnumbers,nofootinbib,superscriptaddress,notitlepage]{revtex4-1}

\usepackage{graphics}
\usepackage{bm}
\usepackage{cancel}
\usepackage{color}
\usepackage{xcolor}
\usepackage{multirow}
\usepackage{slashed}
\usepackage{array}
\usepackage{amssymb}
\usepackage{graphicx}
\usepackage{mathrsfs}
\usepackage{diagbox}
\usepackage{amsmath}
\usepackage{float}
\usepackage{extarrows}
\usepackage[colorlinks,citecolor=blue,anchorcolor=red,menucolor=red,linkcolor=red,filecolor=red,runcolor=red,urlcolor=red,frenchlinks=red]{hyperref}
\usepackage{subfigure}
\usepackage{ulem}

\makeatletter
\newcommand{\thickhline}{\noalign {\ifnum 0=`}\fi \hrule height 1pt\futurelet \reserved@a \@xhline}
\newcolumntype{"}{@{\hskip\tabcolsep\vrule width 1pt\hskip\tabcolsep}}
\makeatother

\def\pslash{\not{\hbox{\kern-4pt $p$}}}
\def\qslash{\not{\hbox{\kern-4pt $q$}}}
\def\lv{\not{\hbox{\kern-4pt $L$}}}
\def\lsim{\mathrel{\raise.3ex\hbox{$<$\kern-.75em\lower1ex\hbox{$\sim$}}}}
\def\gsim{\mathrel{\raise.3ex\hbox{$>$\kern-.75em\lower1ex\hbox{$\sim$}}}}
\def\ifmath#1{\relax\ifmmode #1\else $#1$\fi}

\begin{document}

\title{NLO EW corrections to tau pair production via photon fusion\\ in Pb-Pb ultraperipheral collision}

\author{Jun Jiang}
\email{jiangjun87@sdu.edu.cn}
\affiliation{School of Physics, Shandong University, Jinan, Shandong 250100, China}

\author{Peng-Cheng Lu}
\email{pclu@sdu.edu.cn}
\affiliation{School of Physics, Shandong University, Jinan, Shandong 250100, China}

\author{Zong-Guo Si}
\email{zgsi@sdu.edu.cn}
\affiliation{School of Physics, Shandong University, Jinan, Shandong 250100, China}

\author{Han Zhang}
\email{han.zhang@mail.sdu.edu.cn}
\affiliation{School of Physics, Shandong University, Jinan, Shandong 250100, China}

\author{Xin-Yi Zhang}
\email{xinyizhang@mail.sdu.edu.cn}
\affiliation{School of Physics, Shandong University, Jinan, Shandong 250100, China}

\begin{abstract}

We study the next-to-leading order (NLO) electroweak (EW) corrections to the $\gamma \gamma \to  \tau^+ \tau^-$ process in Pb-Pb ultraperipheral collision (UPC).
We find that the EW correction $\delta \sigma_{\mathrm{EW}}$ decreases the total cross section $\sigma_{\mathrm{NLO}} = \sigma_{\mathrm{LO}} + \delta \sigma_{\mathrm{EW}}$ by -3\% at Pb-Pb center-of-mass energy $\sqrt{s_{NN}}=5.02$ TeV.
The weak correction plays significant role whose contribution is about -4 times of that of QED.
The CMS and ATLAS collaborations use the reaction $\gamma\gamma \to  \tau^+ \tau^-$ in Pb-Pb and proton-proton UPC to constrain tau's anomalous magnetic moment $a_\tau$.
By parameterizing the $\gamma \tau \tau$ vertex with two form factors $F_{1,2}$, the cross section can be written as $\sigma_{a_\tau} = \sigma_{\mathrm{LO}} + \delta \sigma_{a_\tau}$, where $\delta \sigma_{a_\tau}$ is proportional to $a_\tau$.
The impact of NLO EW corrections on $a_\tau$ bounds in Pb-Pb UPC is limited, as the current experimental bounds are loose.
We also find that various differential distributions of the two ratios $\mathrm{d} \sigma_{\mathrm{NLO}}/ \mathrm{d}  \sigma_{\mathrm{LO}}$ and $\mathrm{d} \sigma_{a_\tau}/ \mathrm{d}  \sigma_{\mathrm{LO}}$ have different lineshapes. 
This work is significant to precisely study the interaction of $\gamma \tau \tau$ via $\gamma \gamma \to  \tau^+ \tau^-$ process.

\end{abstract}

\maketitle

\section{INTRODUCTION}\label{sec1}

The Standard Model (SM) of particle physics becomes mature, and the  search for New Physics (NP) becomes one of the hot topics.
In the precise era, the anomalous magnetic moment (AMM) of leptons shall be a smoking gun of NP which lies in the joints of accurate theoretical and experimental studies. 
The value of the electron's AMM predicted by Quantum Electrodynamics (QED) matches the experimental result with remarkable precision of ten significant digits \cite{PhysRevLett.109.111807,doi:10.1126/science.aap7706}, making it the most accurate measurement in nature.
Recently, the muon's g-2 experiment at Fermilab reports the muon's AMM of $a_{\mu}^{\mathrm{exp}}=1.16592059(22) \times 10^{-3}$ \cite{Muong-2:2023cdq},
which agrees with the SM prediction $a_\mu^{\mathrm{SM}} = 1.16591810(43) \times 10^{-3}$ with precision of five significant digits \cite{Aoyama:2020ynm}, showing a deviation of $\Delta a_{\mu } =a_{\mu}^{\mathrm{exp}}- a_{\mu}^{\mathrm{SM}}=0.00000249(48)\times 10^{-3}$ with 5.2$\sigma$
and the latest most precise SM prediction using the data-driven dispersive approach gives the deviation of $\Delta a_{\mu } =a_{\mu}^{\mathrm{SM}} - a_{\mu}^{\mathrm{exp}} = -(123 \pm 49) \times 10^{-11}$ with $2.5 \sigma$ \cite{Davier:2023fpl}.
In 2024, the CMS collaboration observes the tau pair production via photon fusion in proton-proton collision and constrains tau's AMM, $-0.0042 < a_\tau^{\mathrm{CMS}} < 0.0062$ at 95$\%$ C.L. \cite{CMS:2024qjo}, while the SM prediction is $a_\tau^{\mathrm{SM}} = 1.17721(5) \times 10^{-3}$ \cite{Eidelman:2007sb}.
It is worth noting that the measurement of $a_{\tau}$ is challenging due to the tau's short lifetime of approximately $10^{-13}$ seconds, which precludes the use of spin precession method used in the $a_e$ and $a_\mu$ measurements \cite{Muong-2:2006rrc}. 


The $\gamma l^+ l^-$ ($l=e,\ \mu,\ \tau$) coupling vertex has the form of $-i e \Gamma^\mu$ where $\Gamma^\mu$ can be parameterized with the most general form,
\begin{equation} \label{ytt}
    \Gamma^\mu (\frac{q^2}{m^2}) = \gamma^\mu F_1(\frac{q^2}{m^2}) + \frac{\sigma^{\mu\nu } q_\nu}{2m} \left[ i F_2(\frac{q^2}{m^2}) +F_3 (\frac{q^2}{m^2}) \gamma_5 \right],
\end{equation}
where $F_{i}(\frac{q^2}{m^2})$ $(i=1,2,3)$ are form factors which contain the complete electric and magnetic coupling information, $\sigma^{\mu\nu } = i [\gamma^\mu, \gamma^\nu]/2$ is the spin tensor, $q$ is the momentum transfer between outgoing and incoming leptons of mass $m$. 
$F_1(0)=1$,
$F_2 (0) = a_l \equiv (g_l-2)/2$ is the AMM which is a powerful measurement to probe NP,
and $F_3(0)=-2m d_l/e$ with $d_l$ being the anomalous electric dipole moment which is beyond the scope of this paper.

The lepton's AMM $a_{l}$ has two resources: corrections from higher order Feynman diagrams and the compositeness of leptons.
New particles from NP will make their contributions in loop corrections at higher order.
In the supersymmetric model with energy scale $\Lambda_{\mathrm{SUSY}}$, the correction to leptons's AMM scales with the squared lepton mass, $\delta a_{l} \propto \frac{m_l^{2}}{\Lambda_{\mathrm{SUSY}}^{2}}$ \cite{PhysRevD.64.035003}.
Since the leptons' mass ratios are $m_\mu/m_e \approx 207$ and $m_\tau/m_\mu \approx 17$,
we have much greater opportunity to catch a glimpse of NP by the measurement of tau's AMM.

The photon-induced heavy-ion or proton-proton ultraperipheral collision (UPC) at Large Hadron Collider (LHC) can be treated as a photon-photon collider.
An UPC event is that two nuclei collide at a distance greater than twice their radii and always remain intact during the collision.
In UPC, the highly relativistic ions or protons become a strong source of electromagnetic radiation, which can be considered
as fluxes of quasi-real photons in the equivalent photon approximation (EPA) \cite{vonWeizsacker:1934nji,Williams:1934ad}.
A photon-induced UPC event is a pure QED process which has no noisy hadronic background.
In heavy-ion UPC, the photon flux is enhanced by the squared ion charge $Z^2$ for each ion, and the cross section is largely enhanced by an overall $Z^4$ factor.
In addition, photon-induced UPC can provide de-excitation photons with energies up to $\sim$100 GeV \cite{Shao:2022cly}. 

The photon-induced heavy-ion UPC has been used to study the production of lepton pairs and set constrains on the lepton's AMM.
Several experiments have observed the production of electron pairs \cite{PhysRevC.70.031902,PHENIX:2009xtn,ALICE:2013wjo,STAR:2019wlg} and muon pairs \cite{ATLAS:2020epq,CMS:2020skx} via photon-induced heavy-ion UPC. 
For the tau pair production, the ATLAS collaboration observes the $\gamma \gamma \longrightarrow \tau^+ \tau^-$ process in Pb-Pb UPC and sets the constraint $-0.057 <a_{\tau }<0.024$ at 95$\%$ C.L. \cite{ATLAS:2022ryk}, and the CMS also observes the process in Pb-Pb UPC and obtains a model-dependent estimate of $a_\tau = 0.001^{+0.055}_{-0.089}$ \cite{CMS:2022arf}.
Several phenomenological efforts are also made to make constraints on $a_{\tau}$ using the photon-induced heavy-ion UPC \cite{delAguila:1991rm,Beresford:2019gww,Dyndal:2020yen,Shao:2023bga}.
In addition, the NLO QED correction to $\gamma\gamma \longrightarrow l^+ l^- \, (l=\mu,\tau)$ in Pb-Pb UPC is investigated \cite{Shao:2024dmk}, and the NLO EW correction to such process in $e^+e^-$ collision is also discussed \cite{Demirci:2021zwz}.
In this work, we study the next-to-leading order (NLO) electroweak (EW) correction to the cross section of $\gamma\gamma \longrightarrow \tau^+ \tau^-$ in Pb-Pb UPC.
The measurement of $a_\tau$ through the reaction $\gamma\gamma \longrightarrow \tau^+ \tau^-$ is discussed.

The rest of the paper is organized as follows. 
In Section \ref{sec2}, we show the framework of the calculation of tau pair production via photon fusion in Pb-Pb UPC, and introduce the NLO EW corrections to $\gamma\gamma \longrightarrow \tau^+ \tau^-$ process.  
In Section \ref{sec3}, we present the numerical results and some discussion on the measurement of $a_\tau$.
Section \ref{sec4} is reserved for the summary.
In Appendix \ref{appA}, we discuss the feasibility of $F_{1,2}$ parametrization scheme to the $\gamma\tau\tau$ vertex in $\gamma \gamma \longrightarrow \tau^+ \tau ^-$ reaction.

\section{NLO EW correction to $\gamma\gamma \longrightarrow \tau^+ \tau^-$ in Pb-Pb UPC}\label{sec2}
\begin{figure}[!htbp]
    \centering
    \includegraphics[width=0.8\textwidth]{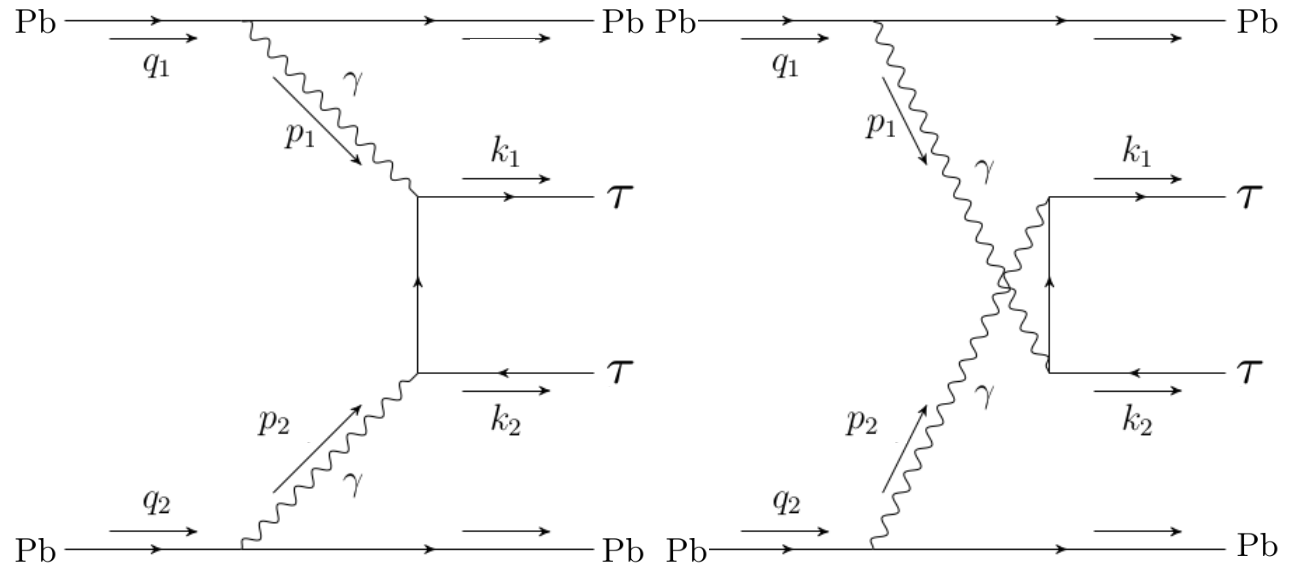}
    \caption{Tau pair production via photon fusion in the Pb-Pb UPC.}
    \label{pb-pb}
\end{figure}

In this section, we show the framework of the calculation for the photon-induced tau pair production in Pb-Pb ultraperipheral collision (UPC) as illustrated in Fig. \ref{pb-pb}, and introduce the NLO EW corrections for the reaction $\gamma\gamma \longrightarrow \tau^+ \tau^-$. 

For an UPC event, the distance between two nuclei, defined as the impact parameter $b$, is larger than the sum of charge radii of two nuclei $A$ and $B$, $b> R_A + R_B$. 
This limits the virtuality $Q^2$ of the photon coherently emitted from a charged nucleus $A$ to be $Q^2 \sim 1/R_A^2 < 4 \times 10^{-3} $ $\mathrm{GeV}^{-2}$ for $ R_A \simeq 1.2 A^{\frac{1}{3}} $ fm with mass number $A>16$ \cite{Shao:2022cly}.
Thus the EPA, also known as the Weizsacker-Williams approximation (WWA)  \cite{vonWeizsacker:1934nji,Williams:1934ad}, is sufficient to describe the radiated photons in the photon-induced processes in UPC.
With such small virtuality, one might use such photon-induced UPC processes to probe the $a_\tau$ since the squared momentum transfer $q^2 = -Q^2 \sim 0$, as what ATLAS \cite{ATLAS:2022ryk} and CMS \cite{CMS:2022arf} have done. 
The cross section for the production of tau pair in photon-induced Pb-Pb UPC
can be formulated as 
\begin{equation}\label{eq:totalXS}
\sigma(\mathrm{Pb} \mathrm{Pb} \stackrel{\gamma \gamma}{\longrightarrow} \mathrm{Pb} \mathrm{Pb} \tau^+ \tau^-) =\int \mathrm{d}x_{1} \mathrm{d}x_{2} n(x_{1}) n(x_{2}) \sigma(\gamma \gamma \to \tau^{+}\tau^{-}),
\end{equation}
with
\begin{equation}\label{n(x)}
\begin{aligned}
n(x_i)=\frac{2Z^{2} \alpha}{x_i \pi } \left \{ \bar{x}_iK_{0}(\bar{x}_i ) K_{1}(\bar{x}_i)-\frac{\bar{x}_i^{2}}{2} [K_{1}^{2}(\bar{x}_i)-K_{0}^{2}(\bar{x}_i) ]  \right \}.
\end{aligned}
\end{equation}
Here, $n(x_i)$ is the effective photon spectrum integrated over impact parameter $b$ \cite{Jackson:1998nia,Shao:2022cly}, charge number $Z=82$  for Pb, $x_{i}=E_{i}/E_{beam}$ is the ratio of the energy $E_{i}$ of the emitted photons from ion $i$ to the beam energy $E_{beam}$, $\bar{x}_i =x_{i} m_{N} b_{min} $,  $m_{N}$ is the mass of the nucleon $m_{N}=0.9315$ GeV, the minimum of impact parameter $b_{min}$ is set to be the nuclear radius $b_{min} =R_{A} \simeq 1.2A^{\frac{1}{3} } \mathrm{fm} = 6.09A^{\frac{1}{3} }$ GeV$^{-1}$ with $A=208$ for Pb, and $K_0,~K_1$ are the modified Bessel functions of the second kind of zero and first order, respectively.
In the UPC, photons emitted by heavy ions are coherently radiated, which imposes a limit on the maximum of center-of-mass (CM) energy of two photons, $\sqrt{s_{\gamma\gamma}^{max} } =\frac{\gamma_L}{R_A}$
where $\gamma_L$ is the Lorentz factor $\gamma_L = E_{beam}/m_N$ \cite{Baur:2001jj}.

Note that, in the framework above, we have set the heavy ions Pb to always remain intact, {\it i.e.}, we ignore the dissociation effect from broken Pb.
Only under this assumption, the initial two-photon distribution can be factorized into two parton-distribution-function-like spectrum $n(x_1)n(x_2)$ in Eq. \eqref{eq:totalXS}.
Otherwise, a two-photon differential distribution of the photon energy has to be embedded into Eq. \eqref{eq:totalXS} rather than $n(x_1)n(x_2)$.
One can refer to Ref. \cite{Shao:2022cly} for more information on the $b$ dependent two-photon distribution.
An explicit example of this dissociation effect in Ref. \cite{Knapen:2016moh} shows that the negative correction may reach up to 20\%.
Thus, our estimations of cross sections should be an upper limit.
However, the photon-induced UPC events can be well identified in experiments because there are no hadronic products other than those from tau pair decay and the unbroken ions have typical rapidity gap with the decay products of tau pair.
In addition, the effects of linear polarization of photon and impact parameter dependence of the azimuthal asymmetry for $\gamma\gamma \longrightarrow l^+l^-$ ($l=e,\mu,\tau$) process in Pb-Pb UPC are discussed in Refs. \cite{Li:2019yzy,Li:2019sin}. 

The differential cross section for $\gamma(p_1) \gamma(p_2) \longrightarrow \tau^+(k_1) \tau^-(k_2)$ process has the simple form
\begin{equation}\label{eq:M2}
\mathrm{d}\sigma(\gamma \gamma \to \tau ^{+}\tau ^{-}) = \frac{\left |M  \right | ^{2} }{2 s_{\gamma \gamma } } \mathrm{d}\Pi_2 ,
\end{equation}
where the squared CM energy of two photons $s_{\gamma \gamma}=(p_1+p_2)^2$, and $\mathrm{d}\Pi_2$ is the two-body phase space measure. The Feynman amplitude $M$ reads,
\begin{equation}\label{eq:M}
\begin{aligned}
M=&(-ie^2)\epsilon _{1\mu } \epsilon _{2\nu }\bar{u}(k_{1} )\\
&\times [i\Gamma ^{\mu }(p_1 )\frac{i(\not{p_{t} } +m )}{p^2_t-m^{2} +i\epsilon } i\Gamma ^{\nu }(p_2 )\\
&+i\Gamma ^{\nu }(p_2 )\frac{i(\not{p_{u} } +m )}{p^2_u-m^{2} +i\epsilon }i\Gamma ^{\mu }(p_1 )]v(k_{2} ).
\end{aligned}
\end{equation}
Here, $\epsilon _{1\mu }$ and  $\epsilon _{2\nu }$ are the polarization vectors of initial photons, $p_ {t}=p_ {2}-k_ {2}=k_ {1}-p_ {1}$ and $p_{u}=p_ {1}-k_ {2}=k_ {1}-p_ {2} $ are momenta of the propagators for the Feynman diagrams in $t$ and $u$ channels, respectively.
The most general parametrization of $\Gamma^\mu$ from $\gamma l^+ l^-$ coupling vertex has been defined in Eq. \ref{ytt}. 
In the calculation, we omit $F_3$, $F_1(0)=1$ and $F_2(0)=a_\tau$ because of the on-shell initial photons, so the $\Gamma^\mu$ has the simple form,
\begin{equation} \label{ytt2}
    \Gamma^\mu = \gamma^\mu + i \frac{\sigma^{\mu\nu } q_\nu}{2m} a_\tau.
\end{equation}
Then the squared amplitudes $|M|^2$ would have five terms which are proportional to $a_\tau$ with powers from zero to four.
The term that is free of $a_\tau$ is defined as the squared amplitudes at leading order (LO),
\begin{equation}\label{eq:LO}
\begin{aligned}
\left | M_{\mathrm{LO}} \right |^2= -\frac{64\pi^2 \alpha^2\left [ \beta^4(y^4-2y^2+2)+2\beta^2(y^2-1)-1 \right ] }{(\beta^2y^2 -1)^2} ,
\end{aligned}
\end{equation}
where $\beta=\sqrt{1-\rho}$ with $\rho=4m^2/s_{\gamma\gamma}$, and $ y = \mathrm{cos}\theta $ with $\theta$ being the scattering angle between $p_1$ and $k_1$ in the CM of two photons.
It is obvious that the squared LO amplitudes has the symmetry under $y \to -y$.


\begin{figure}[!htbp]
	\begin{center}
		\subfigure[]{\label{selfenergy}
			\includegraphics[width=0.18\textwidth]{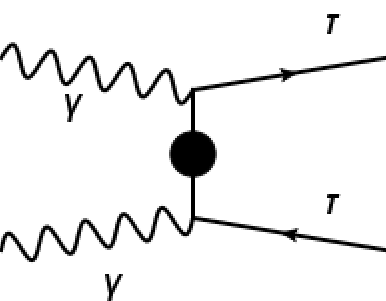} }
		\hspace{-0.3cm}~
             \subfigure[]{\label{vertex}
         	\includegraphics[width=0.18\textwidth]{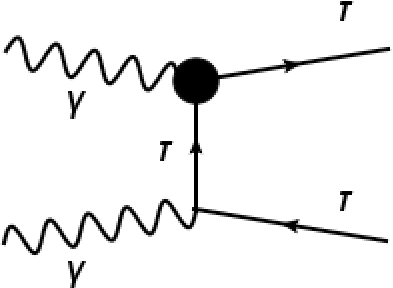} }
		\hspace{-0.3cm}~
		\subfigure[]{\label{AAH}
        	\includegraphics[width=0.2\textwidth]{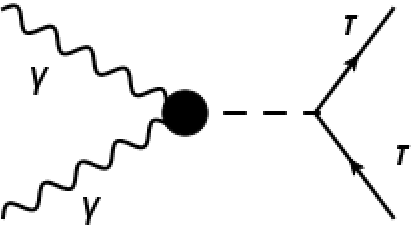} }
        \hspace{-0.3cm}~
	 \\ 
       \subfigure[]{\label{seagull}
        	\includegraphics[width=0.18\textwidth]{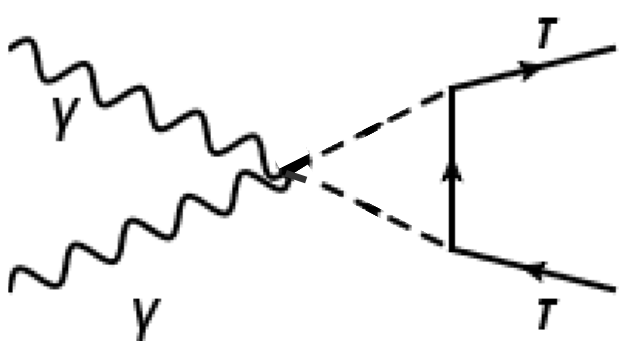} }
		\hspace{-0.3cm}~
       \subfigure[]{\label{box1}
        	\includegraphics[width=0.18\textwidth]{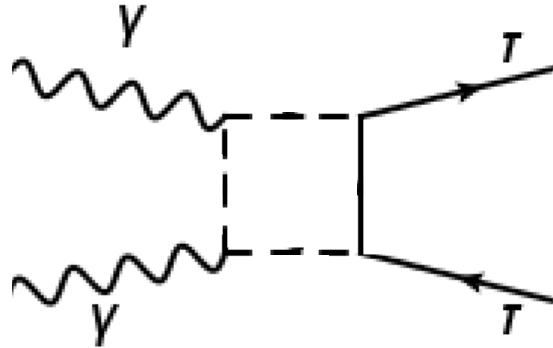} }
       \hspace{-0.3cm}~
       \subfigure[]{\label{box2}
	        \includegraphics[width=0.18\textwidth]{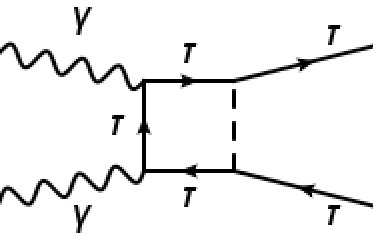} }
        \hspace{-0.3cm}~
        \subfigure[]{\label{real}
        	\includegraphics[width=0.18\textwidth]{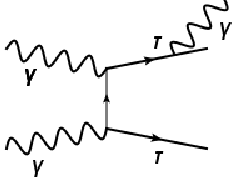} }
		\caption{Typical Feynman diagrams of NLO EW correction to $\gamma \gamma \to \tau ^{+}\tau ^{-} $ process. The black solid spots contain all possible one-loop EW interactions including the counter terms, and the dashed lines might be scalar Higgs or Goldstone bosons, or vector bosons accordingly.}\label{oneloop}
	\end{center}
\end{figure}

The NLO EW correction to the $\gamma \gamma \to \tau^+ \tau^-$ process has typical Feynman diagrams as exhibited in Fig. \ref{oneloop}.
The Feynman diagram can be divided into six types based on their topological structures: self-energy diagrams like Fig. \ref{selfenergy}, 
vertex  correction diagrams with corrections to $\gamma\tau \tau$ vertex like Fig. \ref{vertex}, 
triangle diagrams which contain contributions from massive fermion or boson loops like Fig. \ref{AAH}, 
seagull diagrams with four-particle-coupling correction like Fig. \ref{seagull},
box diagrams like Fig. \ref{box1} and Fig. \ref{box2},
and the real correction diagrams like Fig. \ref{real}.
Since we adopt the Feynman-$'t$ Hooft gauge, Feynman diagrams with the Goldstone bosons are considered.
In these figures, black solid spots contain all possible one-loop EW corrections including the counter terms, and
dashed lines might be scalar Higgs or Goldstone bosons, or vector bosons accordingly.

The cross section with NLO EW corrections to $\gamma\gamma \longrightarrow \tau^+{\tau^-}$ process can be factorized into the following form,
\begin{equation}
\label{C0function}
\sigma(\gamma \gamma \to \tau ^{+}\tau ^{-})=\frac{\alpha ^2}{m^2}\left \{c^{(0)}(\eta)+4 \pi\alpha\left [ c^{(1)}_{\mathrm{QED}}(\eta)+c^{(1)}_{\mathrm{weak}}(\eta)\right ] \right \},
\end{equation}
which is a function of dimensionless variable $\eta=1/\rho -1$. 
$c^{(0)}(\eta)$ is the contribution at LO which is proportional to $|M_{\mathrm{LO}}|^2$ in Eq. \ref{eq:LO}, $c^{(1)}_{\mathrm{QED}}(\eta)$ is the contribution from pure QED correction where the real correction is included,
and $c^{(1)}_{\mathrm{weak}}(\eta)$ stands for the contribution of weak corrections. Note that all $c$ functions are dimensionless.

During the calculation of EW corrections, one encounters the problem of $\gamma_5$ in dimensional regularization. 
We use the strategy by Chen in Refs. \cite{Chen:2023lus,Chen:2024zju} to handle this issue, and we further check the results using the $\gamma_5$ scheme proposed by Kreimer in Ref. \cite{kreimer1994rolegamma5dimensionalregularization}.
Very recently, a program $g5anchor$ is released to help you to deal with the $\gamma_5$ problem \cite{Chen:2024zju}.

Another problem one has to handle is the ultraviolet (UV) divergence and infrared (IR) divergences at NLO calculation.
We use dimensional regularization with $D=4-2\varepsilon$ to regularize both the UV and IR divergences. 
Following the techniques for one-loop EW corrections in Ref. \cite{Denner:1991kt}, the on-shell renormalization scheme is adopted to eliminate the UV divergences.
The IR divergences in loop correction shall be canceled by those in the real correction.
And the IR divergence in real correction is proportional to the cross section at LO \cite{Harris:2001sx},
\begin{equation}\label{eq:IR}
    d\sigma^{\mathrm{IR}}_{\mathrm{real}} =  \frac{\alpha}{2 \pi} \frac{1}{\varepsilon}\left( 2-\frac{1+\beta^2}{\beta} \mathrm{ln}\frac{1+\beta}{1-\beta} \right) \times d \sigma_{\mathrm{LO}}.
\end{equation}
Once the divergences cancel each other, we obtain the divergence-free cross section of tau pair production in photon-induced Pb-Pb UPC,
\begin{equation}\label{eq:sigNLO}
    \sigma_{\mathrm{NLO}} = \sigma_{\mathrm{LO}} + \delta \sigma_{\mathrm{QED}} +\delta \sigma_{\mathrm{weak}},
\end{equation}
where $\delta \sigma_{\mathrm{QED}}$ consists of the one-loop and real QED corrections, and $\delta \sigma_{\mathrm{weak}}$ are contributions from weak one-loop correction.
Sometimes, the last two terms have a compact form $\delta \sigma_{\mathrm{EW}}$.

In addition, we utilize FeynArts \cite{Hahn:2000kx} to generate NLO EW Feynman amplitudes and simplify them with FeynCalc \cite{shtabovenko2023feyncalc10multiloopintegrals,Mertig:1990an}. 
We adopt Fortran codes written by ourselves to deal with the multi-dimension integrals over the phase spaces of both loop and real corrections.

\section{ Numerical results and analysis}\label{sec3}
In this section, we present our numerical results of the NLO EW corrections to the $\gamma\gamma \longrightarrow \tau^+ \tau^-$ reaction in the Pb-Pb UPC, and then discuss the measurement of $a_\tau$ through this reaction.

For the input parameters adopted, the fine structure constant $\alpha = \sqrt{2} G_{\mu} s^2_W m^2_W /\pi$, where the sine of the weak mixing angle $s^2_W = 1- m^2_W/m^2_Z$, $m_W$ and $m_Z$ are the masses of W boson and Z boson respectively. 
The values of input parameters are
\begin{equation}
\begin{aligned}
\label{9}
& m=1.777 \ \mathrm{GeV}   ~~~~~~ m_W=80.369\ \mathrm{GeV}   ~~~~~~ m_Z=91.188 \ \mathrm{GeV}   ~~~~~~ m_H = 125.20\ \mathrm{GeV} \\
& m_t=172.4 \ \mathrm{GeV}   ~~~~~~ m_b=4.183 \ \mathrm{GeV}    ~~~~~~ m_c=1.273\ \mathrm{GeV} ~~~~~~ G_{\mu}=1.166379\times 10^{-5} \ \mathrm{GeV}^{-2}.
\end{aligned}
\end{equation}

\subsection{NLO EW corrections}

\begin{figure}[!htbp]
	\centering
	\includegraphics[width=0.8\linewidth]{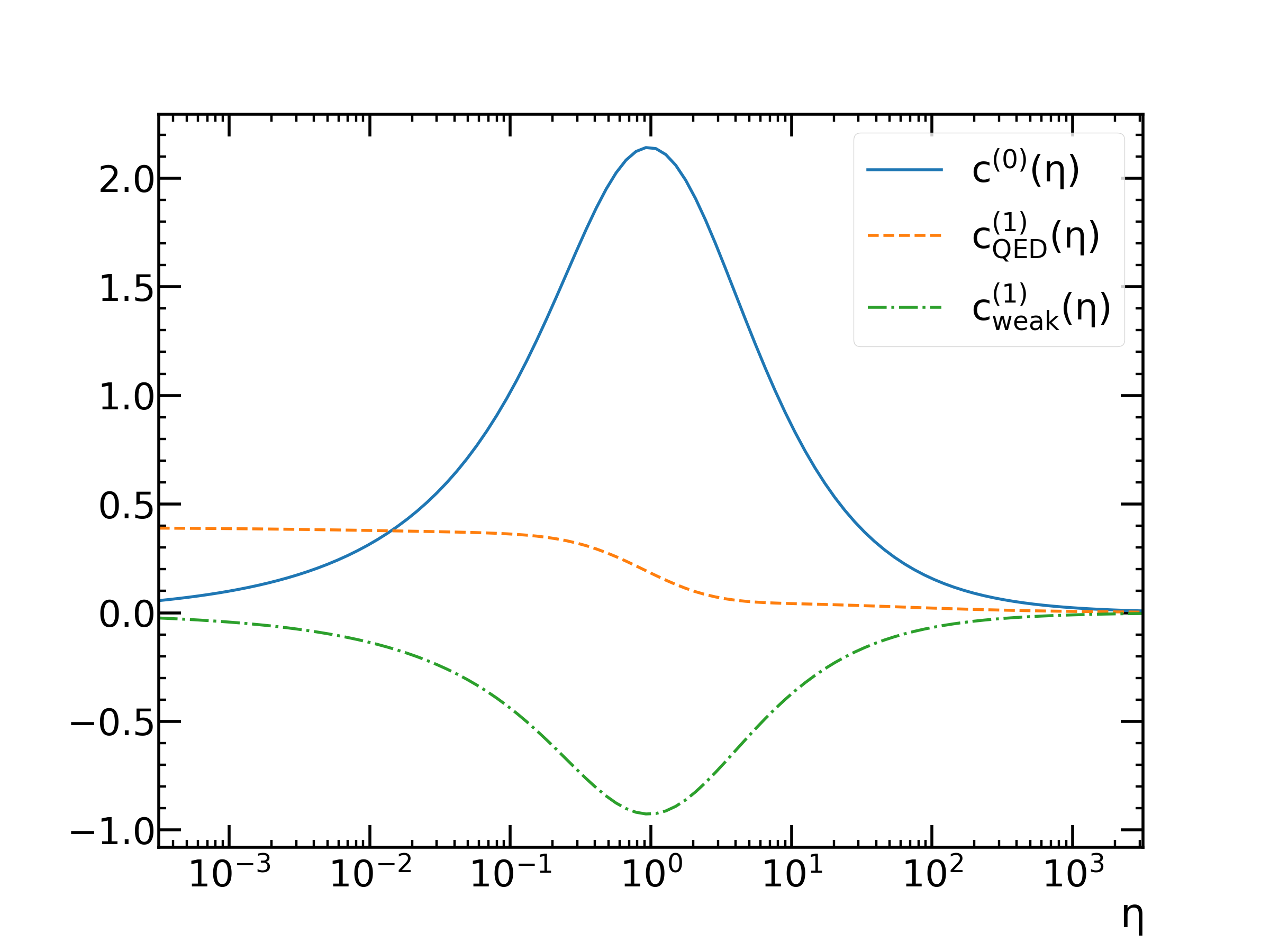}
	\caption{The functions $c^{(0)}(\eta)$, $c^{(1)}_{\mathrm{QED}} (\eta)$, $c^{(1)}_{\mathrm{weak}} (\eta)$ for $\gamma\gamma \longrightarrow \tau^+ \tau^-$ process as functions of the variable $\eta=1/\rho -1$ with $\rho=4m^2/s_{\gamma\gamma}$.}\label{ctoeta}
\end{figure}

We first discuss the NLO EW corrections to the $\gamma\gamma \longrightarrow \tau^+ \tau^-$ process.
The dimensionless functions $c^{(0)}(\eta)$, $c^{(1)}_{\mathrm{QED}}(\eta)$, $c^{(1)}_{\mathrm{weak}}(\eta)$ in Eq. \ref{C0function} as functions of the variable $\eta=1/\rho -1$ with $\rho=4m^2/s_{\gamma\gamma}$ are displayed in Fig. \ref{ctoeta}.
It shows the dependence of contributions of LO, QED and weak corrections on the two-photon CM energy $\sqrt{s_{\gamma \gamma}}$ explicitly. 
In the plot, $c^{(0)}$ has the lineshape of a bump as $\eta$ grows and approaches its maximum 2.1 near $\eta$= 0.92 or $\sqrt{s_{\gamma \gamma}}=4.9$ GeV, while $c^{(1)}_{\mathrm{weak}}$ has a lineshape of a dip as $\eta$ grows and reaches its minimum -0.93 at the same position. 
And $c^{(1)}_{\mathrm{QED}}$ decreases gradually as $\eta$ increases and has a relatively large downgrade around $\eta=1$ or $\sqrt{s_{\gamma \gamma}}=5.0$ GeV.
It is found that total EW correction reaches its maximum which contributes about -3.7$\%$ correction at $\eta$=4.7 or $\sqrt{s_{\gamma \gamma}}=8.5$ GeV.
All three $c$ functions approach zero at large $\eta$ or $\sqrt{s_{\gamma \gamma}}$.

\begin{figure}[!htbp]
	\centering
	\includegraphics[width=0.8\linewidth]{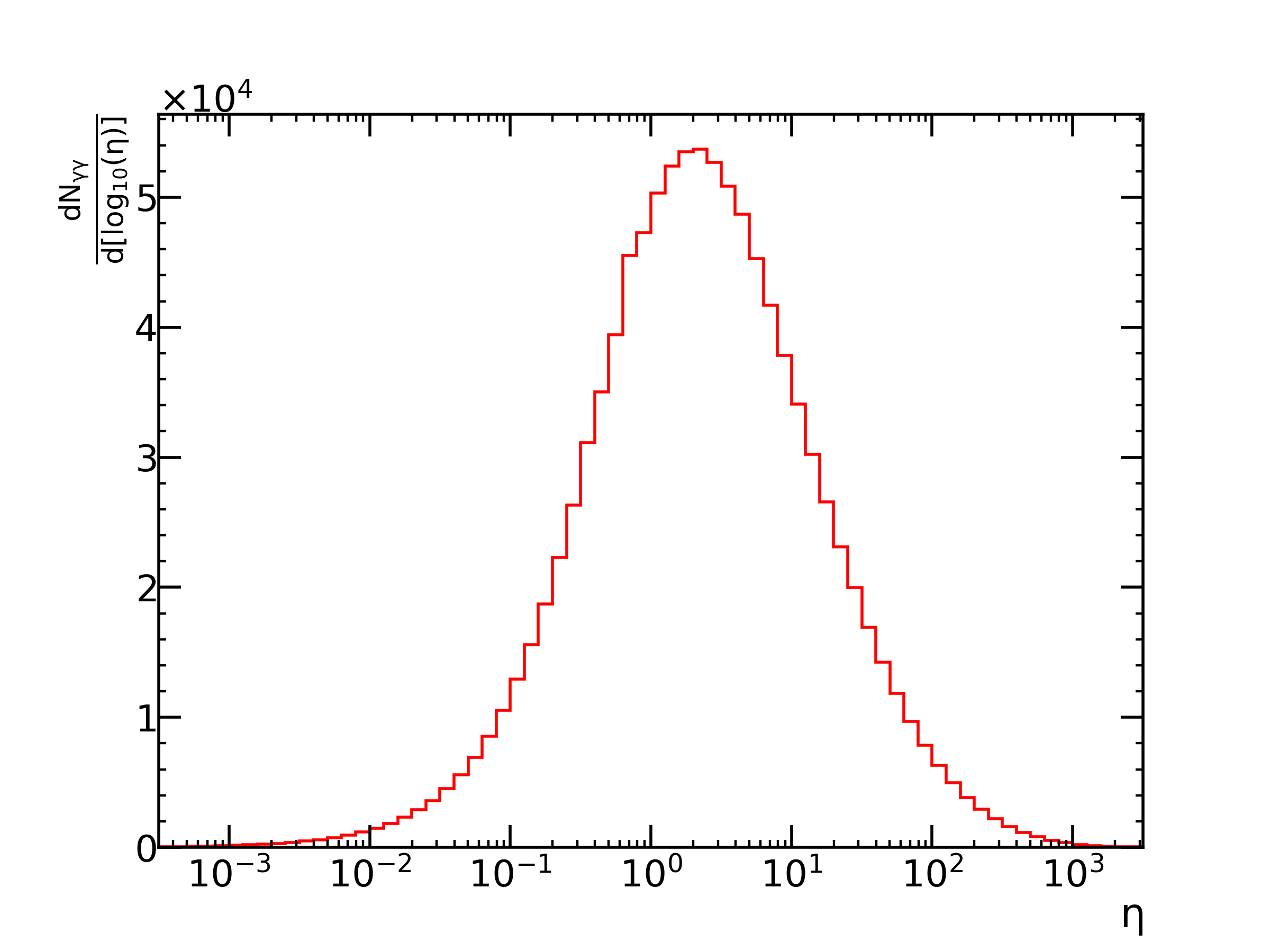}
	\caption{The differential two-photon spectrum $\mathrm{d}N_{\gamma\gamma}/\mathrm{d log_{10}}(\eta)$ as a function of $\eta$.}
	\label{fig:wrr}
\end{figure}

Let's turn to discuss the complete process of tau pair production in photon-induced Pb-Pb UPC.
In Table \ref{table1}, we show the cross sections at different nucleon-nucleon CM energy $\sqrt{s_{NN}}$, where 5.02, 5.36 and 5.52 TeV are available at LHC \cite{Bruce_2020,d_Enterria_2023}.
The $\sigma_{\mathrm{LO}}$, $\delta \sigma_{\mathrm{QED}}$, $\delta \sigma_{\mathrm{weak}}$, and $\sigma_{\mathrm{NLO}}$ are cross sections for LO, QED correction, weak correction, and their sum, respectively.
The total cross section at NLO is of the order of magnitude of about 1 mb.
The total EW correction $\delta \sigma_{\mathrm{EW}} = \delta \sigma_{\mathrm{QED}} + \delta \sigma_{\mathrm{weak}}$ contribute about $-3\%$ to $\sigma_{\mathrm{LO}}$ at typical nucleon-nucleon CM energy at LHC.
In detail, $\delta \sigma_{\mathrm{QED}}$ has about 1\% positive correction to $\sigma_{\mathrm{LO}}$, while $\delta \sigma_{\mathrm{weak}}$ has about 4\% negative correction to $\sigma_{\mathrm{LO}}$.
Each contribution grows with the increase of $\sqrt{s_{NN}}$.
The reason why weak correction $\delta \sigma_{\mathrm{weak}}$ dominates the total NLO EW corrections is that the two-photon spectrum $n(x_1) n(x_2)$ reaches its maximum near the position $\eta=4.7$ where the EW correction to $\gamma\gamma \longrightarrow \tau^+ \tau^-$ process has its negative maximum value.
To explain this, we show the differential two-photon spectrum $\mathrm{d}N_{\gamma\gamma}/\mathrm{d log_{10}}(\eta)$ as a function of $\eta$ in Fig. \ref{fig:wrr}, which is defined as
\begin{equation}\label{eq:Naa}
   \frac{ \mathrm{d}N_{\gamma\gamma} }{\mathrm{d log_{10}}(\eta)} = \int^1_{\frac{4  m^2 (1+\eta)}{s_{NN}}} \frac{4 \mathrm{ln}(10) m^2 \eta}{x_1 s_{NN}}  n\left(x_1 \right) n\left( \frac{4  m^2 (1+\eta)}{x_1 s_{NN}}\right) \mathrm{d}x_1.
\end{equation}
Here, the variable $x_1$ is integrated and $x_2$ has been transformed into $\eta$.
The two-photon spectrum reaches its maximum at the bin of $\eta \in [2.0,2.5]$.

\begin{table}
	\centering
	\begin{tabular}{c c c c c c}
		\hline
		$\sqrt{s_{NN}}$ [TeV] & $\sigma_{\mathrm{LO}}$[mb] & $\delta \sigma_{\mathrm{QED}}$[mb]  & $\delta \sigma_{\mathrm{weak}}$[mb]  & $\sigma_{\mathrm{NLO}}$[mb]  & $\dfrac{\sigma_{\mathrm{NLO}}}{\sigma_{\mathrm{LO}}}$ \\
		\hline
		5.02 & 1.02 & 1.00$\times 10^{-2}$& -4.18$\times 10^{-2}$ & 0.99 & 0.97 \\
		5.36 & 1.08 & 1.06$\times 10^{-2}$ & -4.44$\times 10^{-2}$ & 1.05 & 0.97 \\
		5.52 & 1.11 & 1.09$\times 10^{-2}$ & -4.57$\times 10^{-2}$ & 1.08 & 0.97 \\
		6.00 & 1.20 & 1.17$\times 10^{-2}$ & -4.92$\times 10^{-2}$ & 1.16 & 0.97 \\
		\hline
	\end{tabular}
	\caption{The cross sections of tau pair production via photon fusion in Pb-Pb UPC at different nucleon-nucleon CM energy $\sqrt{s_{NN}}$.}
	\label{table1}
\end{table} 

\begin{figure}[!htbp]
	\begin{center}
		\subfigure[]{\label{NLO Invariant mass distribution}
			\includegraphics[width=0.49\textwidth]{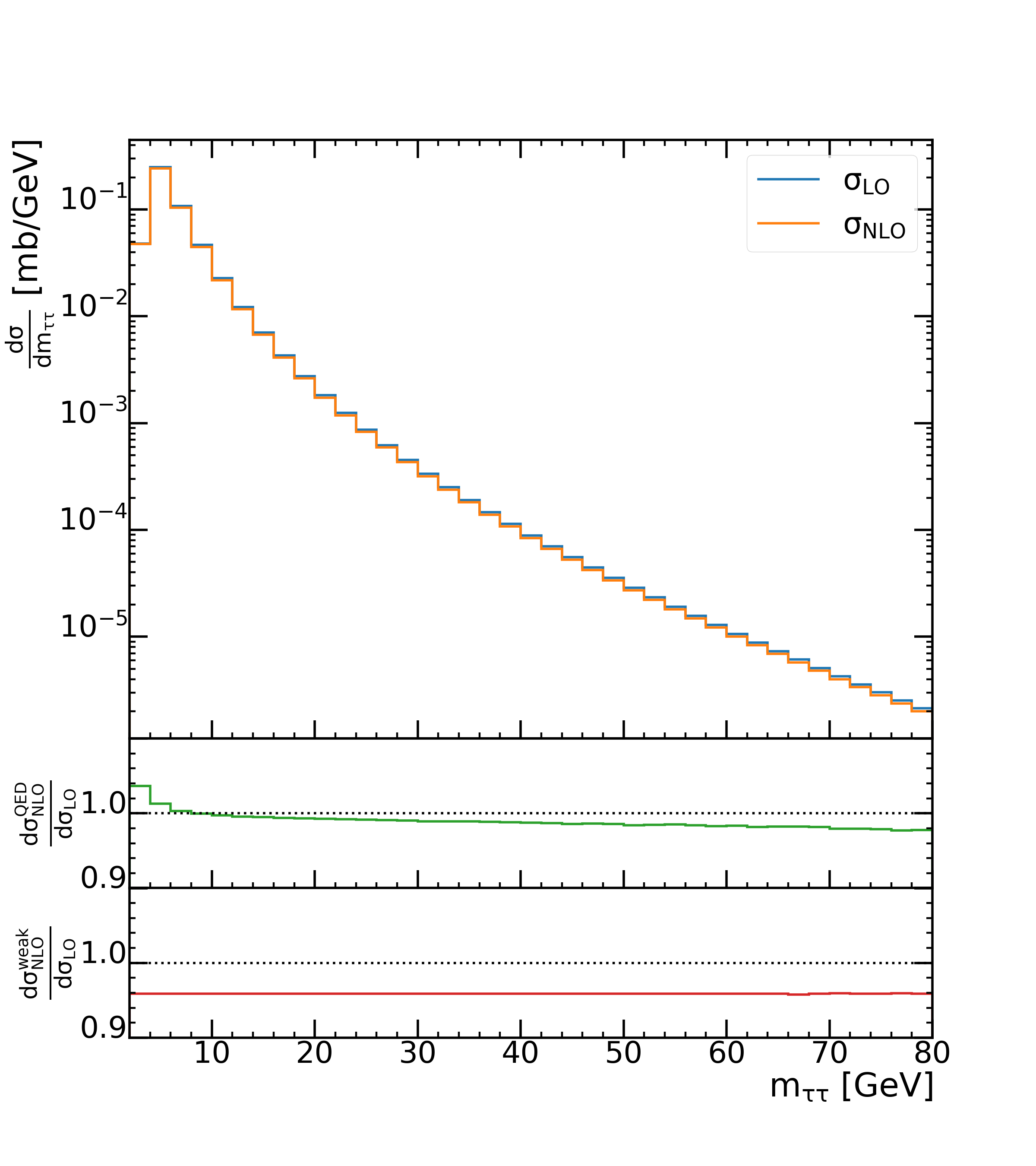} }
		\hspace{-0.9cm}~
		\subfigure[]{\label{NLO pt distribution}
			\includegraphics[width=0.49\textwidth]{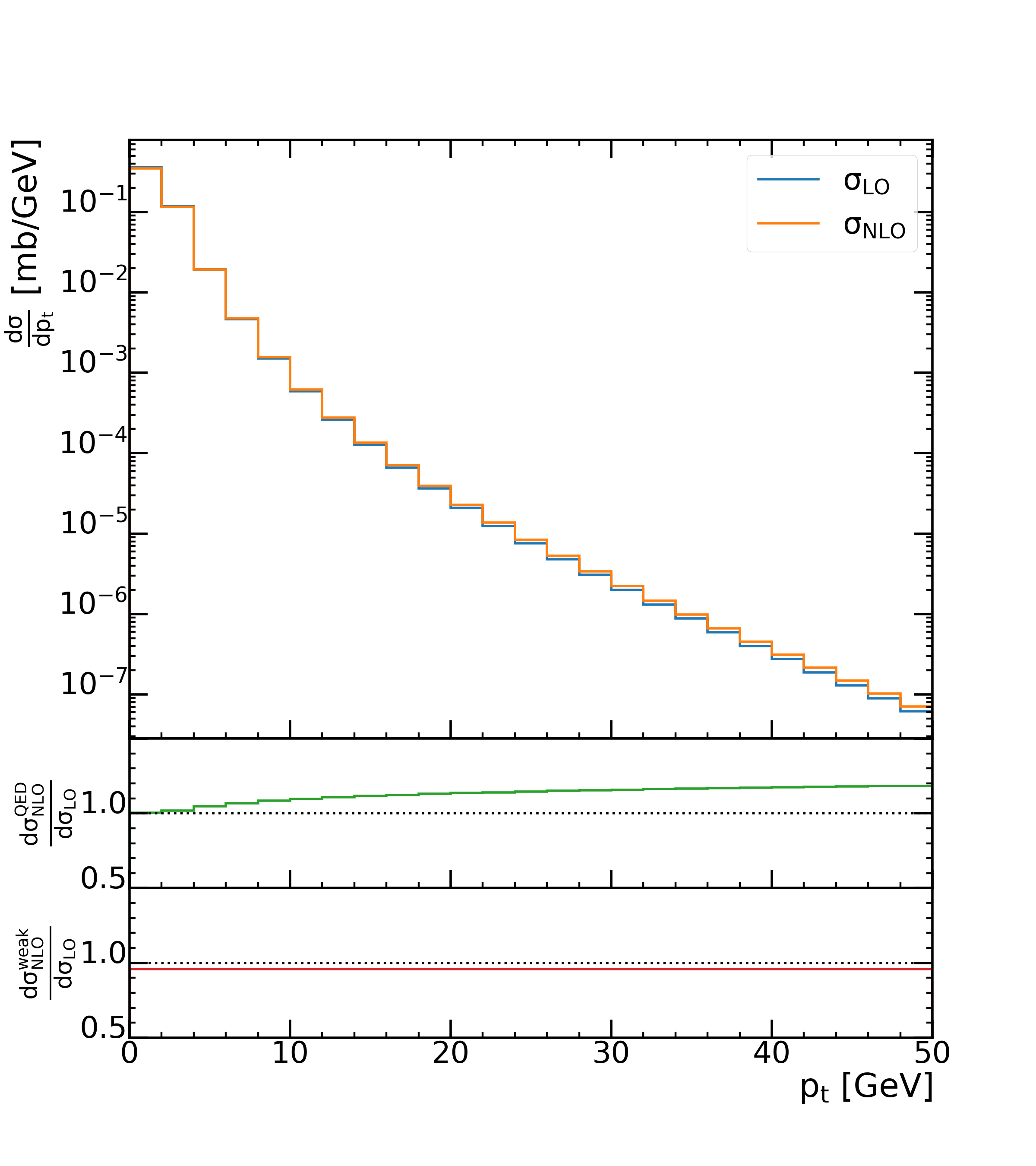} }
		\caption{Differential distributions of cross sections and their ratios to $\sigma_{\mathrm{LO}}$ at $\sqrt{s_{NN}}$=5.02 TeV. Plots in (a) and (b) are for the differential invariant mass distribution of tau pair and the $\tau^-$ transverse momentum distribution, respectively. The up panels are for the NLO and LO cross sections, the central and bottom panels are for the ratios of $\sigma^{\mathrm{QED}}_{\mathrm{NLO}}$ and $\sigma^{\mathrm{weak}}_{\mathrm{NLO}}$ to $\sigma_{\mathrm{LO}}$. }\label{4mass&rapid}
	\end{center}
\end{figure}

In Fig. \ref{4mass&rapid}, 
we show the differential distributions of cross sections and their ratios to $\sigma_{\mathrm{LO}}$ at $\sqrt{s_{NN}}$=5.02 TeV.
The Fig. \ref{NLO Invariant mass distribution} is for the differential invariant mass distribution of tau pair, where the up panel is for the distributions of $\sigma_{\mathrm{LO}}$ and $\sigma_{\mathrm{NLO}}$, and the central and bottom panels are for the ratios of $\sigma^{\mathrm{QED}}_{\mathrm{NLO}}$ and $\sigma^{\mathrm{weak}}_{\mathrm{NLO}}$ to $\sigma_{\mathrm{LO}}$ respectively.
It is found that invariant mass distributions of cross sections for both LO and NLO increase first and then decrease with the increasement of $m_{\tau\tau}$, and the maximum is obtained at the bin of $m_{\tau\tau} \in [4,6]$ GeV.
For the ratio of QED correction, we find that $\mathrm{d} \sigma^{\mathrm{QED}}_{\mathrm{NLO}}/ \mathrm{d}  \sigma_{\mathrm{LO}}$ decreases with $m_{\tau \tau}$ and drop below 1 around $m_{\tau \tau} = $ 8 GeV.
While for the ratio of weak correction, $\mathrm{d} \sigma^{\mathrm{weak}}_{\mathrm{NLO}}/ \mathrm{d}  \sigma_{\mathrm{LO}}$ against $m_{\tau \tau}$ is always below 1 and is almost flat.
This is reasonable as the ratio of $c^{(1)}_{\mathrm{weak}} (\eta)$/$c^{(0)}(\eta)$ is almost a constant in Fig. \ref{ctoeta}.
The QED correction $\sigma^{\mathrm{QED}}_{\mathrm{NLO}}$ contains the contribution of hard real photon radiation which is always positive, and the contribution of virtual correction.
For the real correction, the small $m_{\tau \tau}$ region corresponds to the large energy of radiated photon where the positive real correction dominates. As the $m_{\tau \tau}$ increases, the energy of radiated photon decreases, and the positive contribution of real correction becomes smaller. 
In addition, the virtual correction has a negative contribution in large $m_{\tau \tau}$ region.
These lead to the phenomenon that $\mathrm{d} \sigma^{\mathrm{QED}}_{\mathrm{NLO}}/ \mathrm{d}  \sigma_{\mathrm{LO}}$ decreases with $m_{\tau \tau}$ and drops below 1.
Fig. \ref{NLO pt distribution} is for the differential transverse momentum distribution of $\tau^-$ (similar for $\tau^+$). 
The up panel shows the cross sections decrease monotonously as $\tau^-$'s $p_t$ grows.
The ratio of QED correction to LO cross section  increases as $\tau^-$'s $p_t$ grows, while the ratio of weak correction to LO cross section has a flat lineshape.
The reason why $\mathrm{d} \sigma^{\mathrm{QED}}_{\mathrm{NLO}}/ \mathrm{d} \sigma_{\mathrm{LO}}$ increases with $p_t$ is the same as that why it deceases with $m_{\tau\tau}$.
The small $p_t$ region of tau corresponds to the small $p_t$ region of radiated photon. As the $p_t$ grows, the positive contribution of real correction becomes more and more significant, so the ratio $\mathrm{d} \sigma^{\mathrm{QED}}_{\mathrm{NLO}}/ \mathrm{d} \sigma_{\mathrm{LO}}$ increases with $p_t$.

\subsection{Measurement of $a_\tau$ in $\gamma\gamma \longrightarrow \tau^+ \tau^-$ reaction}

In this subsection, we discuss the measurement of $a_{\tau}$ in the reaction $\gamma\gamma \longrightarrow \tau^+ \tau^-$ under the  $F_{1,2}$ parametrization scheme.
By parameterizing the $\gamma \tau \tau$ vertex with two form factors $F_{1,2}$, there will be five terms which are proportional to $F_2$ with powers from zero to four in the squared amplitudes.
In the on-shell condition $q^2 \to 0$ of initial photons, $F_2(0) = a_\tau$. 
The term proportional to $a_{\tau}$ has the explicit form
\begin{equation}\label{eq:atau}
\begin{aligned}
\left | M_{a_{\tau}} \right |^2= -\frac{256\pi^2\alpha^2}{\beta^2y^2-1} a_{\tau}.
\end{aligned}
\end{equation}
In this $F_{1,2}$ parametrization scheme, terms with higher powers of $a_\tau$ are ignored, and the total cross section of tau pair production in photon-induced Pb-Pb UPC can be written into two terms,
\begin{equation}\label{eq:sigatau}
\sigma_{a_\tau} = \sigma_{\mathrm{LO}} + \delta \sigma_{a_\tau},
\end{equation}
where their differential cross sections $\mathrm{d}\sigma_{\mathrm{LO}} \propto \left| M_{\mathrm{LO}} \right |^2$, and $\mathrm{d}\delta \sigma_{a_\tau} \propto \left| M_{a_{\tau}} \right |^2$.
It is worth emphasizing that $\sigma_{a_\tau}$ has the different definition from $\sigma_{\mathrm{NLO}}$ in Eq. \ref{eq:sigNLO}.
The corrections from NLO and higher order and NP effects would make their contributions to the $\delta \sigma_{a_\tau}$ in Eq. \ref{eq:sigatau}.

In experimental measurements using Pb-Pb UPC, ATLAS \cite{ATLAS:2022ryk} extracts the value of $a_{\tau}$ by performing a template fit to the muon transverse-momentum distribution from tau lepton candidates, while CMS  \cite{CMS:2022arf} use the dependency of the total cross section $\sigma(\gamma\gamma \longrightarrow \tau^+ \tau^-)$ as a function of $a_\tau$ to extract a model-dependent \cite{Beresford:2019gww}
value of $a_\tau$.
It is worth noting that, only the vertex correction diagrams like Fig. \ref{vertex} will contribute to $a_\tau$ in theoretical view.
For example, the LO contribution $a_\tau = \alpha/2\pi$ comes from the triangle Feynman loop that a photon is exchanged between the tau pair in the $\gamma \tau \tau$ vertex.
However, in experiments, all higher order Feynman diagrams in SM and the NP effects, including irrelevant Feynman diagrams like Figs. \ref{selfenergy}, \ref{AAH}, \ref{seagull}, \ref{box1} and Fig. \ref{box2}, will make their contributions to the experimental measurement of $a_\tau$.

\begin{figure}[!htbp]
	\centering
	\includegraphics[width=0.9\textwidth]{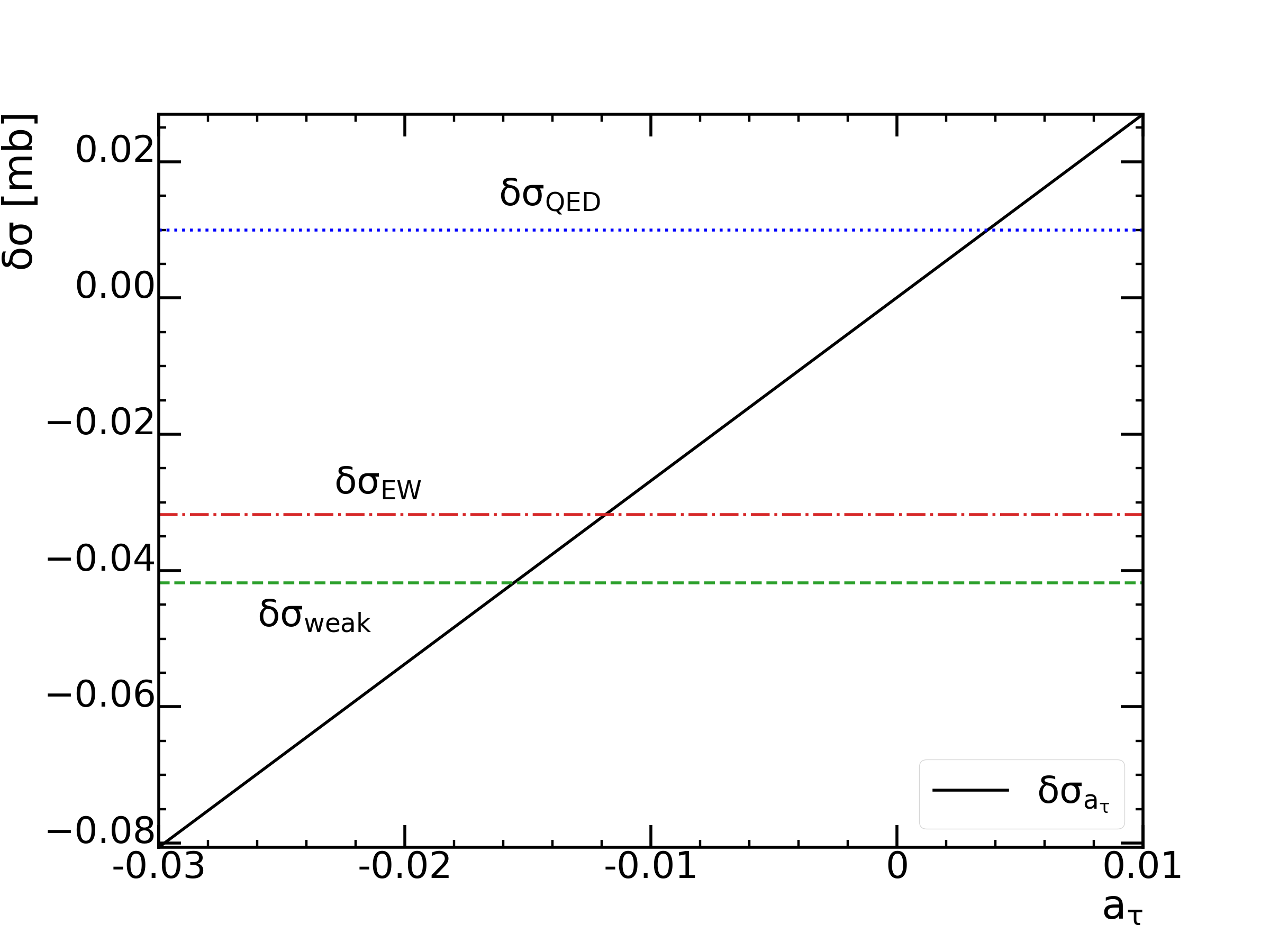}	
	\caption{The correction $\delta \sigma_{a_\tau} $ as a function of $a_\tau$ and the values of NLO EW corrections at $\sqrt{s_{NN}} = 5.02$ TeV. $\delta \sigma_{a_\tau}$ stands for the correction defined in Eq. \ref{eq:sigatau} in $F_{1,2}$ parametrization scheme, which is proportional to $a_\tau$. $\delta \sigma_{\mathrm{QED}} = 1.00 \times 10^{-2}$ mb, $\delta \sigma_{\mathrm{weak}} = -4.18 \times 10^{-2}$ mb and their sum $\delta \sigma_{\mathrm{EW}} = -3.18 \times 10^{-2}$ mb at $\sqrt{s_{NN}} = 5.02$ TeV.}
	\label{ataus}
\end{figure}

To have a more intuitive image how $a_\tau$ is related to the corrections to cross sections, we show the correction $\delta \sigma_{a_\tau} $ as a function of $a_\tau$ and the values of NLO EW corrections at $\sqrt{s_{NN}} = 5.02$ TeV in Fig. \ref{ataus}.
In the plot, $\delta \sigma_{a_\tau}$ is the correction defined in Eq. \ref{eq:sigatau} in the $F_{1,2}$ parametrization scheme, which is proportional to $a_\tau$.
And $\delta \sigma_{\mathrm{QED}}$, $\delta \sigma_{\mathrm{weak}}$ and $\delta \sigma_{\mathrm{EW}}$ are the corrections from QED, weak and their sum at NLO, which are $1.00 \times 10^{-2}$ mb, $-4.18 \times 10^{-2}$ mb and $-3.18 \times 10^{-2}$ mb at $\sqrt{s_{NN}}=5.02$ TeV, respectively.
The ATLAS collaboration has observed the $\gamma \gamma \longrightarrow \tau^+ \tau^-$ process in Pb-Pb UPC and sets the constraint $-0.057 <a_{\tau }<0.024$ at 95$\%$ C.L. \cite{ATLAS:2022ryk} based on the differential transverse-momentum fit, and the CMS conducts the same measurement and obtains a model-dependent \cite{Beresford:2019gww} estimate of $a_\tau = 0.001^{+0.055}_{-0.089}$ \cite{CMS:2022arf} based on the total inclusive UPC cross section under the $F_{1,2}$ parametrization scheme. 
As the current bounds are loose, our estimate is helpful for the precise measurement of $a_\tau$ using Pb-Pb UPC in future.
Note that, the most latest and more stringent constraint on $a_\tau$ is $-0.0042 < a_\tau^{\mathrm{CMS}} < 0.0062$ at 95$\%$ C.L. by CMS collaboration in 2024, which however is extracted from p-p UPC using the $\gamma \gamma \longrightarrow \tau^+ \tau ^-$ reaction \cite{CMS:2024qjo}. 
This measurement is based on differential cross section analyses at high $m_{\tau \tau} > 50$ GeV region, and differs significantly from Pb-Pb UPC in phase space.
Other measurements of $a_\tau$ by DELPHI in 2004 using $\gamma \gamma \longrightarrow \tau^+ \tau ^-$ reaction in $e^+e^-$ collision \cite{DELPHI:2003nah}, and by OPAL \cite{OPAL:1998dsa} and L3 \cite{L3:1998gov} in 1998 using $e^+e^- \longrightarrow Z^0 \longrightarrow \tau^+ \tau ^- \gamma$ reaction, have large uncertainties at 95\% C.L., all of which are largely beyond the range we plot.

\begin{figure}[!htbp]
	\begin{center}
		\subfigure[]{\label{Invariant mass distribution}
			\includegraphics[width=0.34\textwidth]{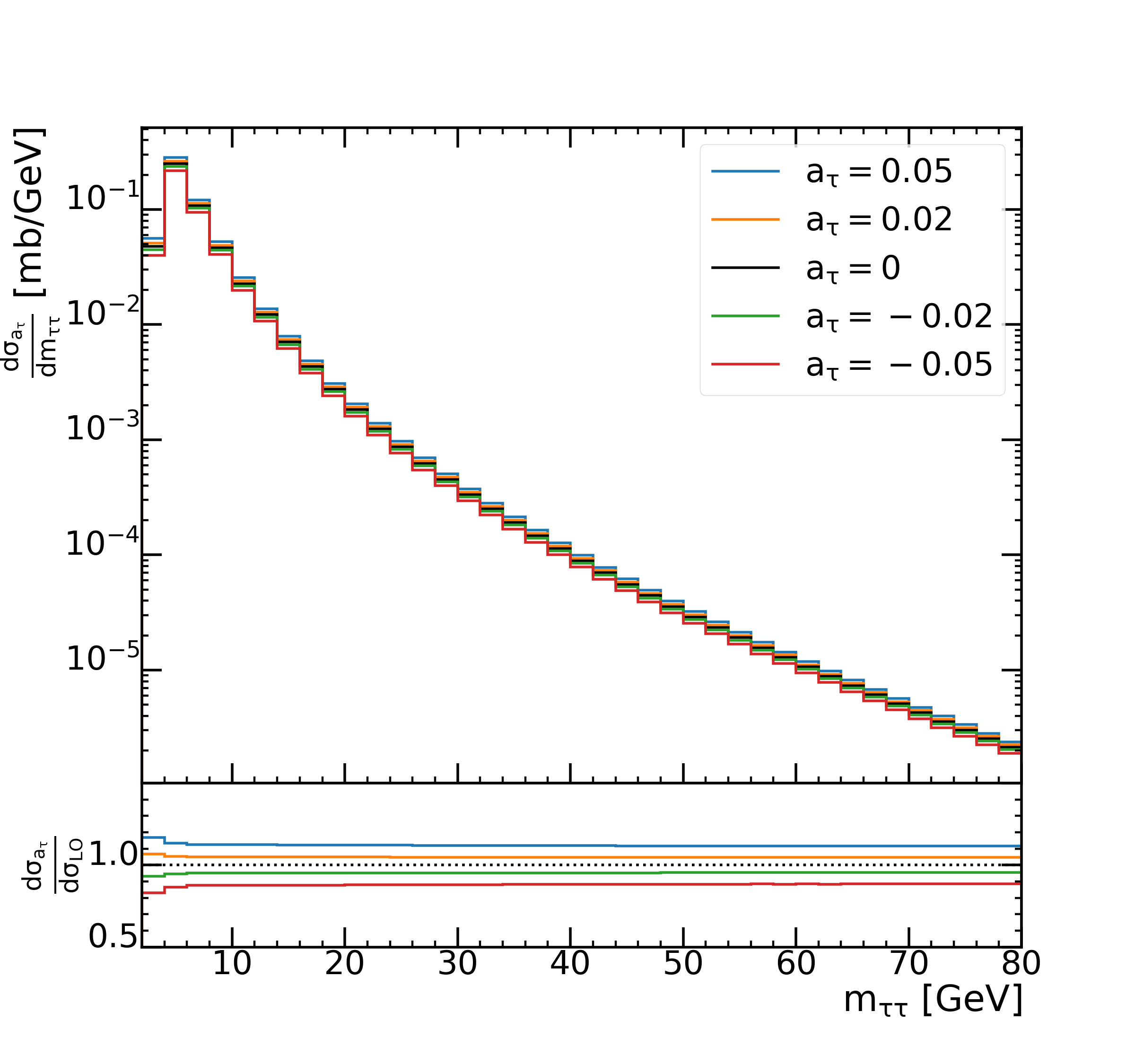} }
		\hspace{-0.9cm}~
		\subfigure[]{\label{pt distribution}
			\includegraphics[width=0.34\textwidth]{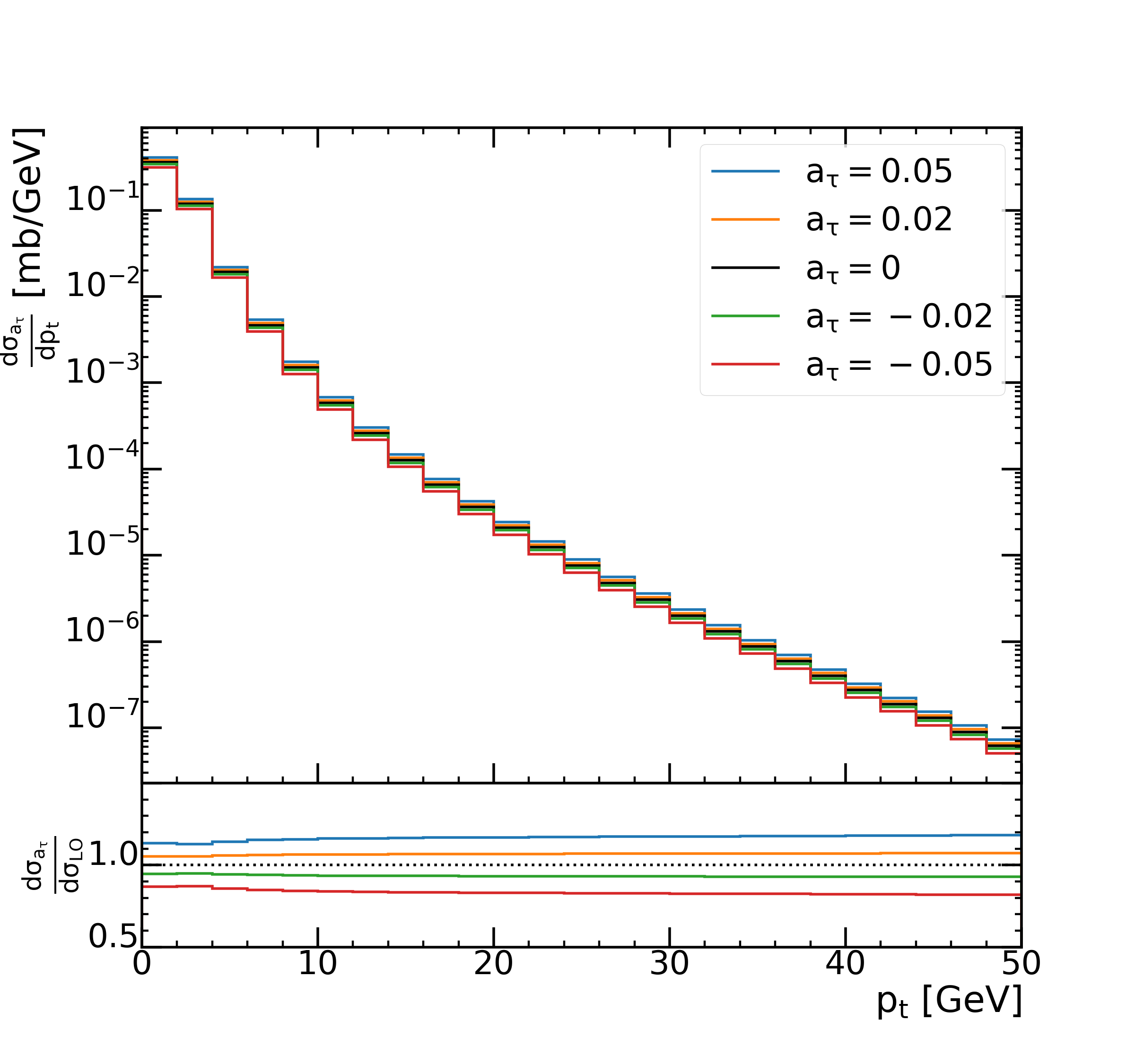} }
   	\hspace{-0.9cm}~
		\subfigure[]{\label{Rapidity distribution}
		\includegraphics[width=0.34\textwidth]{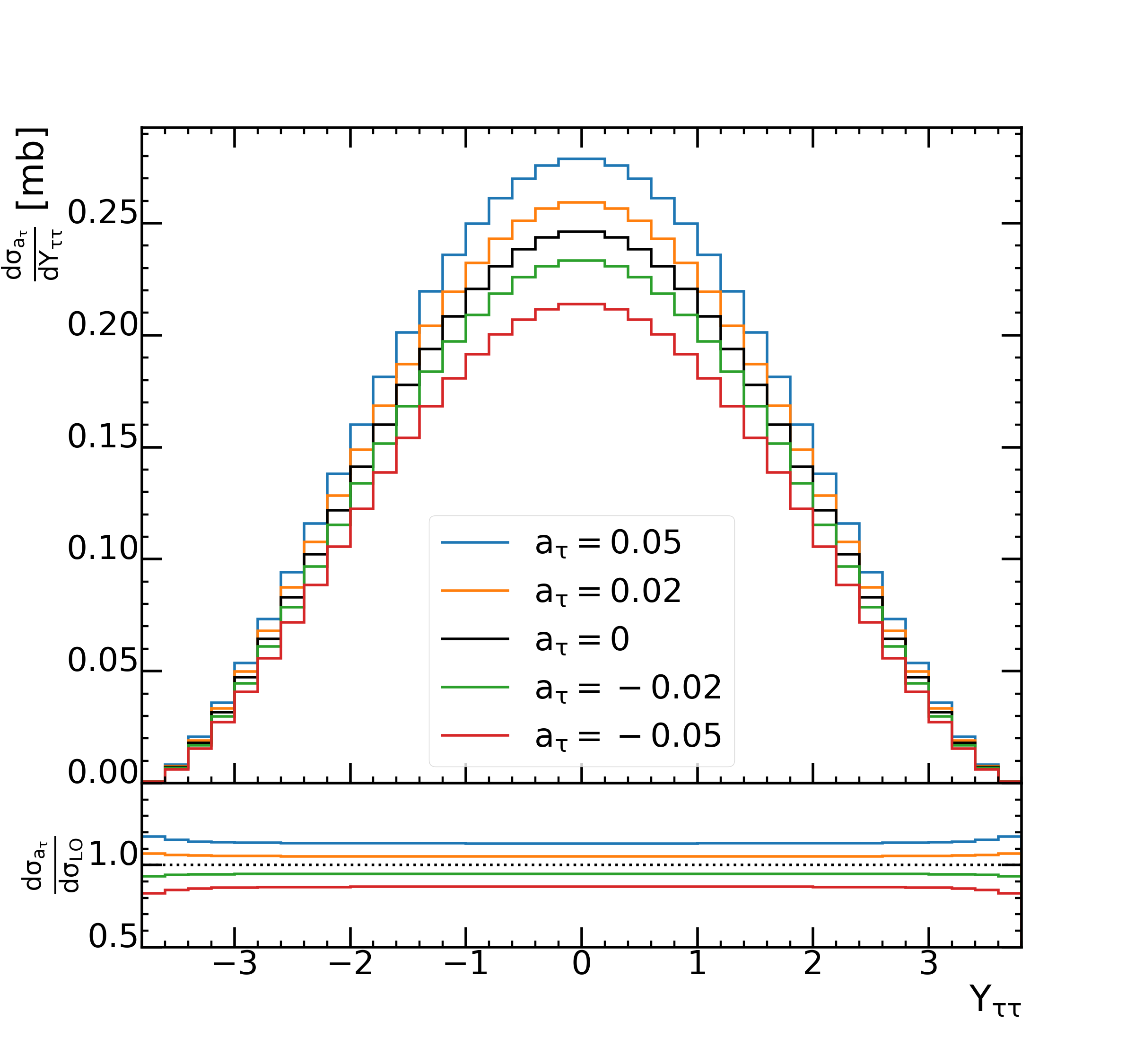} }
    	\caption{Differential distributions of the cross section $\sigma_{a_\tau}$ (up panels), and its ratio to $\sigma_{\mathrm{LO}}$ (down panels) at $\sqrt{s_{NN}}$=5.02 TeV. Plots in (a), (b) and (c) are for the differential invariant mass distribution of tau pair, the $\tau^-$ transverse momentum distribution, and the rapidity distribution of tau pair, respectively.}\label{5mass&rapid}
	\end{center}
\end{figure}

In Fig. \ref{5mass&rapid}, we show various differential distributions of cross sections $\sigma_{a_\tau}$ and its ratio to $\sigma_{\mathrm{LO}}$ at $\sqrt{s_{NN}}$=5.02 TeV.
The up  and bottom panels are for $\sigma_{a_\tau}$ and the ratio $\mathrm{d}\sigma_{a_\tau}/\mathrm{d}\sigma_{\mathrm{LO}}$, respectively.
And Fig. \ref{Invariant mass distribution}, Fig. \ref{pt distribution}, and Fig. \ref{Rapidity distribution} are for the differential invariant mass distribution of tau pair, the $\tau^-$ transverse momentum distribution, and the rapidity distribution where $Y_{\tau\tau} = \frac{1}{2} \mathrm{ln} \frac{E_{\tau\tau} +P_{\tau\tau,z}}{E_{\tau\tau} -P_{\tau\tau,z}}$ with tau pair energy $E_{\tau\tau}$ and momentum $P_{\tau\tau,z}$ in $z$ direction of laboratory frame, respectively.
To illustrate the influence of $a_\tau$ on the differential parametrized cross section $\mathrm{d} \sigma_{a_\tau}$, we adopt five different values $a_\tau = 0,\ \pm0.02, \ \pm0.05$. 
The invariant mass distribution $\frac{\mathrm{d} \sigma_{a_\tau}}{\mathrm{d} m_{\tau\tau}} $ increases first and then decreases with the increase of $m_{\tau\tau}$, and the maximum is obtained at the bin of $m_{\tau\tau} \in [4,6]$ GeV.
Note that $m_{\tau\tau} = \sqrt{s_{\gamma\gamma}}$ for $\sigma_{a_\tau}$, but not true for $\sigma_{\mathrm{NLO}}$ since the later contains the real correction.
The transverse momentum distribution $\frac{\mathrm{d} \sigma_{a_\tau}}{\mathrm{d} p_t}$ decreases monotonously as $\tau^-$'s $p_t$ grows.
The rapidity distribution $\frac{\mathrm{d} \sigma_{a_\tau}}{\mathrm{d} Y_{\tau\tau}} $ is much more sensitive to $a_\tau$ at the central rapidity region than at large rapidity region, while it is the opposite for its ratio $\mathrm{d} \sigma_{a_\tau}/\mathrm{d} \sigma_{\mathrm{LO}}$.

By comparing the differential distributions in Fig. \ref{4mass&rapid} and Fig. \ref{5mass&rapid}, we find that distributions of the ratio $\mathrm{d} \sigma_{\mathrm{NLO}}/ \mathrm{d}  \sigma_{\mathrm{LO}}$ have different lineshapes in comparison with those of $\mathrm{d} \sigma_{a_\tau}/ \mathrm{d}  \sigma_{\mathrm{LO}}$.
This implies that extracting the $a_\tau$ value by fitting the differential distributions of $\gamma \gamma \longrightarrow \tau^+ \tau^-$ reaction in Pb-Pb UPC under the $F_{1,2}$ parametrization scheme might not be a good idea.
Two reasons should be responsible for such discrepancies.
First, there exist further contributions in addition to corrections to $\gamma \tau \tau$ vertex, whose typical Feynman diagrams are shown in Figs. \ref{selfenergy}, \ref{AAH}, \ref{seagull}, \ref{box1} and Fig. \ref{box2}.
Second, there exist extra form factors other than $F_{1,2}$ in the one-loop correction to the $\gamma \tau \tau$ vertex in  reaction $\gamma \gamma \to  \tau^+ \tau^-$.
We will discuss those new form factors in detail, {\it i.e.} the feasibility of $F_{1,2}$ parametrization scheme in $\gamma\gamma \longrightarrow \tau^+ \tau^-$ in Appendix \ref{appA}.

\section{Summary}\label{sec4}
We study the NLO EW corrections to the $\gamma\gamma \longrightarrow \tau^+ \tau^-$ process in the Pb-Pb UPC and discuss the measurement of $a_{\tau}$ in this reaction.
The NLO corrections from QED and weak interactions are illustrated in Fig. \ref{ctoeta} as functions of dimensionless variable $\eta = 1/\rho -1$ with $\rho = 4m^2/s_{\gamma\gamma}$.
The cross sections of tau pair production via photon fusion in Pb-Pb UPC at different nucleon-nucleon CM energy $\sqrt{s_{NN}}$ are displayed in Table \ref{table1}.
We show differential distributions with respect to the invariant mass $m_{\tau\tau}$, and the transverse momentum $p_t$ of $\tau^-$ for cross sections $\sigma_{\mathrm{NLO}}$, and its ratios to $\sigma_{\mathrm{LO}}$ at $\sqrt{s_{NN}}$=5.02 TeV in Fig. \ref{4mass&rapid}.
By parameterizing the $\gamma \tau \tau$ vertex with two form factors $F_{1,2}$, the cross section can be written as $\sigma_{a_\tau} = \sigma_{\mathrm{LO}} + \delta \sigma_{a_\tau}$, where $\delta \sigma_{a_\tau}$ is proportional to $a_\tau$.
We present $\delta \sigma_{a_\tau}$ as a function of $a_\tau$, and make comparison with NLO corrections in Fig. \ref{ataus}. 
The differential distributions with respect to invariant mass $m_{\tau\tau}$, the transverse momentum $p_t$ and the rapidity $Y_{\tau\tau}$ for the cross section $\sigma_{a_\tau}$ and its ratio to $\sigma_{\mathrm{LO}}$ are shown in Fig. \ref{5mass&rapid}.
Hints on the measurement of $a_\tau$ from the differential cross sections are discussed.

We find that the NLO QED correction $\delta \sigma_{\mathrm{QED}}$ is positive, while the weak correction $\delta \sigma_{\mathrm{weak}}$ is negative and is about a factor of minus four of $\delta \sigma_{\mathrm{QED}}$.
The weak correction plays significant role in the NLO correction to $\gamma\gamma \longrightarrow \tau^+ \tau^-$ reaction.
The total EW correction contributes about $-3\%$ to the LO result at typical nucleon-nucleon CM energy at LHC.
We find that the NLO EW correction $\delta \sigma_{\mathrm{EW}}$ does not match either the range implied by the CMS constraint on $a_\tau$ or the value derived from SM prediction of $a_\tau$ under the $F_{1,2}$ parametrization scheme.
The impact of NLO EW corrections on $a_\tau$ bounds in Pb-Pb UPC is limited, as the current bounds are loose.
Our results are helpful for the precise measurement of $a_\tau$ using Pb-Pb UPC in future.
It is also found that various differential distributions of the two ratios $\mathrm{d} \sigma_{\mathrm{NLO}}/ \mathrm{d}  \sigma_{\mathrm{LO}}$ and $\mathrm{d} \sigma_{a_\tau}/ \mathrm{d}  \sigma_{\mathrm{LO}}$ have different lineshapes.
Two reasons shall be responsible for those discrepancies.
First, there are further contributions in addition to corrections to $\gamma \tau \tau$ vertex which have to be taken into consideration in the NLO calculation of $\gamma\gamma \longrightarrow \tau^+ \tau^-$.
Second, the NLO EW corrections to $\gamma\tau\tau$ vertex in this reaction show that there are contributions from new form factors other than $F_{1,2}$. 
Due to the much shorter life of tau lepton in comparison with electrons and muons, we can not continue to use the spin precession method to measure $a_\tau$. 
The $\gamma\gamma \longrightarrow \tau^+ \tau^-$ reaction becomes a preferred process.
Our study will be helpful to study the interaction of $\gamma\tau\tau$ via the $\gamma\gamma \longrightarrow \tau^+ \tau^-$ reaction.

\section*{Acknowledgements}

The authors thank Prof. H. F. Li for the valuable discussions, and extend appreciation to the members of the Institute of theoretical physics of Shandong University for their helpful communications. This work is supported in part by National Natural Science Foundation of China under the Grants No. 12235008, No. 12321005, No. 12475083, and No.12405121.

\begin{appendix}
	
	\section{Feasibility of $F_{1,2}$ parametrization scheme for $\gamma\tau\tau$ vertex \\ in $\gamma\gamma \longrightarrow \tau^+ \tau^-$ reaction} \label{appA}
	
	For the relevant process $l^- (k_1) \gamma^*(q) \longrightarrow l^-(k_2)$ with $l=e,\ \mu, \ \tau$, we have the unconstrained off-shell photon and the on-shell external leptons.
	Its matrix element is $  (-ie \epsilon_\mu)  \bar{u}(k_2) \Gamma^\mu(\frac{q^2}{m^2}) u(k_1)$, where
	\begin{equation}
	\Gamma^\mu (\frac{q^2}{m^2}) = 
	\begin{minipage}[b]{0.15\columnwidth}
	\centering
	\raisebox{-.5\height}{\includegraphics[width=1\linewidth]{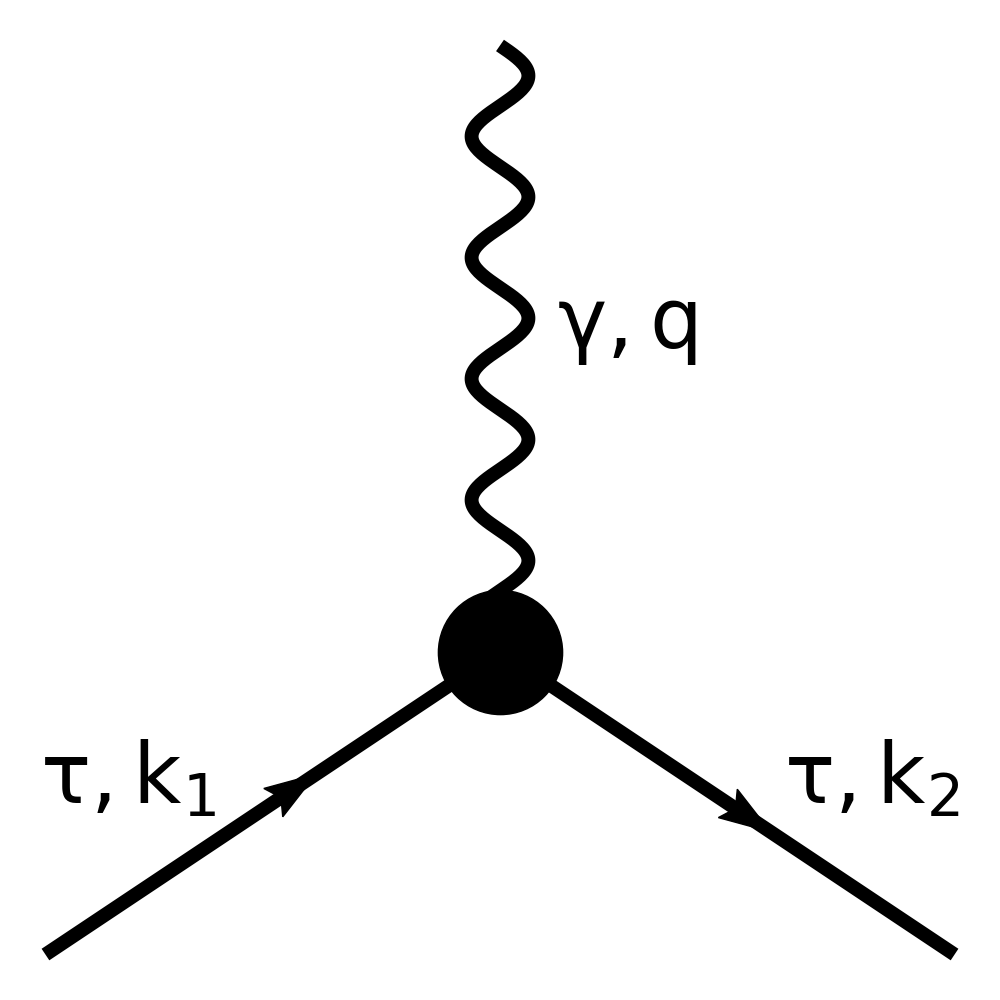}}
	\end{minipage}
	=  \gamma^\mu F_1(\frac{q^2}{m^2}) + i \frac{\sigma^{\mu\nu } q_\nu}{2m} F_2(\frac{q^2}{m^2}).
	\end{equation}
	To derive the above parametrization form, the Gordon identity that works only for on-shell spinors is used.
	This $F_{1,2}$ parametrization scheme for process $l^- (k_1) \gamma^*(q) \longrightarrow l^-(k_2)$ holds to all orders in perturbation theory.
	In such parametrization scheme to $\gamma\tau\tau$ vertex, we calculate the cross section $\sigma_{a_\tau}$ for the $\gamma\gamma \longrightarrow \tau^+ \tau^-$ reaction in Pb-Pb UPC, which are presented in Eqs. \ref{eq:M}-\ref{eq:LO}, \ref{eq:atau} and \ref{eq:sigatau}. 
	
	However, in the calculation of NLO EW correction to $\gamma\gamma \longrightarrow \tau^+ \tau^-$ reaction, we find that the various one-loop weak corrections to $\gamma\tau\tau$ vertex with one tau off-shell have extra Lorentz structures other than $\gamma^\mu$ and $\sigma^{\mu\nu } q_\nu$, in other words, new form factors other than $F_{1,2}$ arise.
	There are two $\gamma \tau \tau$ vertices in reaction $\gamma\gamma \longrightarrow \tau^+ \tau^-$. 
	The typical one-loop weak corrections to $\Gamma^\mu$ in the $\gamma\tau\tau$ vertex in $t$ channel have the following form,
	\begin{align}
	\Gamma_{\mathrm{weak}}^\mu  &= 
	\begin{minipage}[b]{0.15\columnwidth}
	\centering
	\raisebox{-.5\height}{\includegraphics[width=1\linewidth]{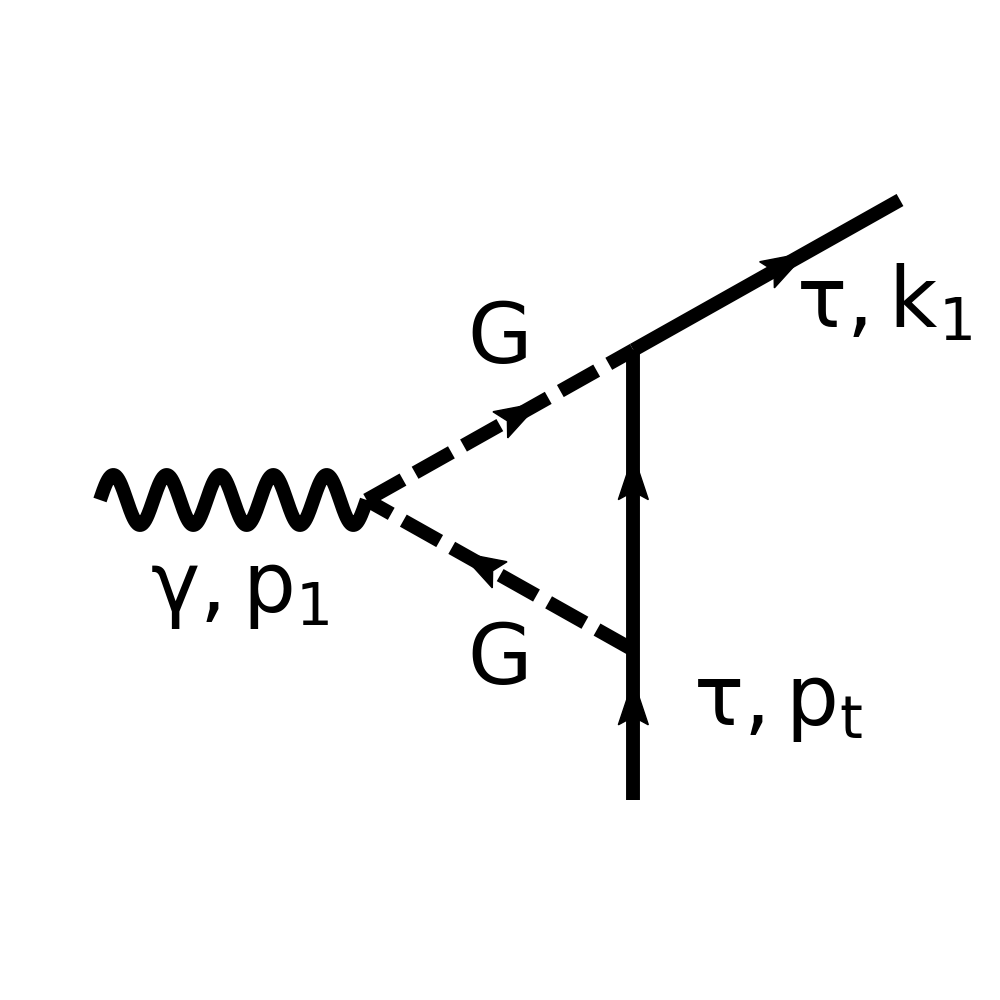}}
	\end{minipage}
	+    \begin{minipage}[b]{0.15\columnwidth}
	\centering
	\raisebox{-.5\height}{\includegraphics[width=1\linewidth]{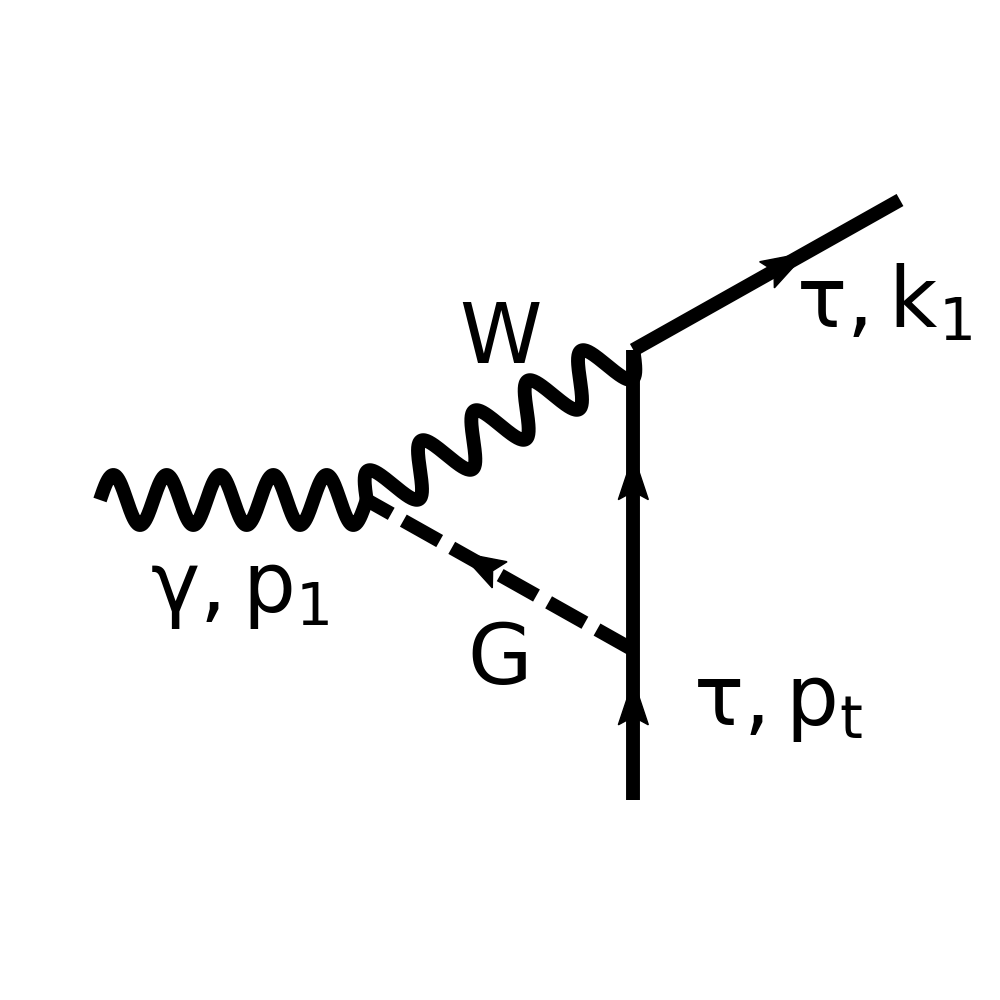}}
	\end{minipage}
	+      \begin{minipage}[b]{0.15\columnwidth}
	\centering
	\raisebox{-.5\height}{\includegraphics[width=1\linewidth]{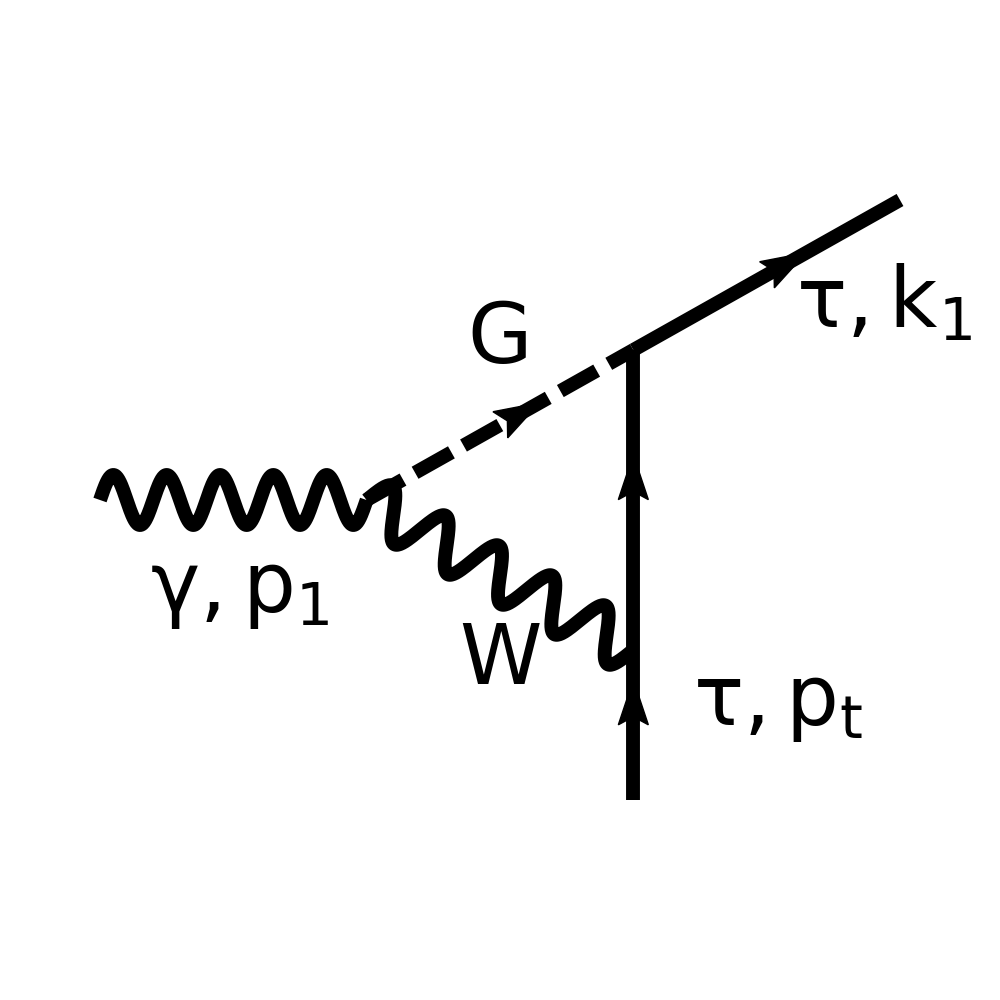}}
	\end{minipage} 
	+ \begin{minipage}[b]{0.15\columnwidth}
	\centering
	\raisebox{-.5\height}{\includegraphics[width=1\linewidth]{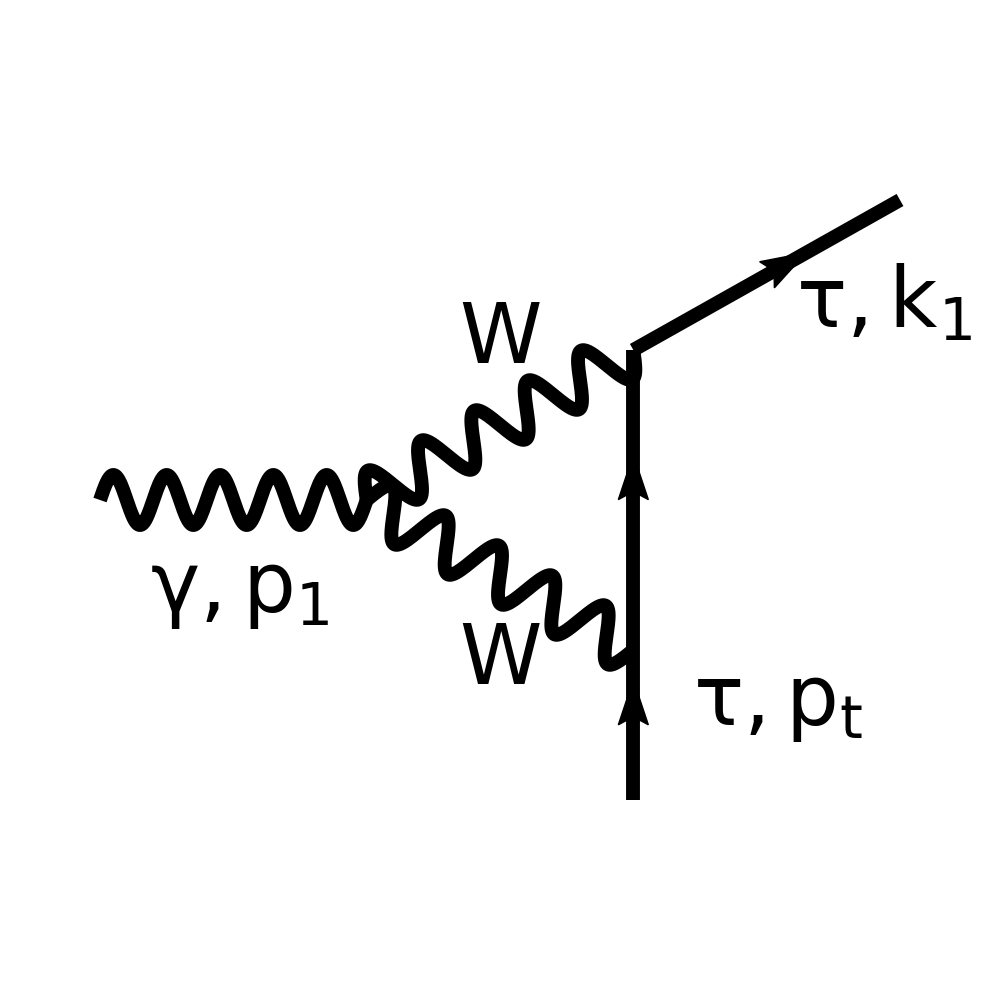}}
	\end{minipage} 
	+ \begin{minipage}[b]{0.15\columnwidth}
	\centering
	\raisebox{-.5\height}{\includegraphics[width=1\linewidth]{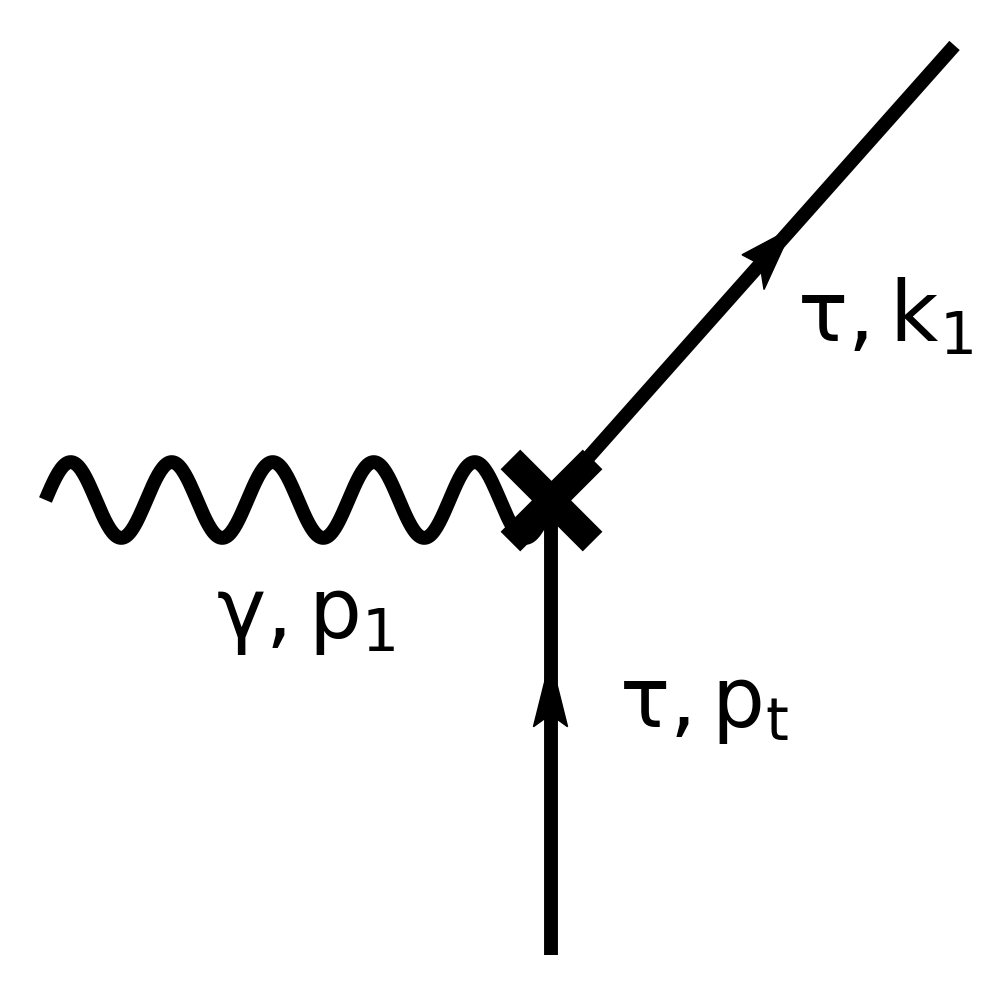}}
	\end{minipage} \nonumber \\
	&= \frac{\alpha}{4 \pi s_W^2} \left(\gamma^{\mu}\widetilde{F}_1 +\gamma^{\mu}\gamma^5 \widetilde{F}_2 +\gamma^{\mu}\not{p_1}\widetilde{F}_3 +\gamma^{\mu}\not{p_1}\gamma^5 \widetilde{F}_4  +k_{1\mu}\not{p_1}\widetilde{F}_5  \right. \nonumber \\&\left. \hspace{1.5cm} +k_{1\mu}\not{p_1}\gamma^5\widetilde{F}_6 +i\frac{\sigma^{\mu\nu}k_{1\nu}}{2m}\widetilde{F}_7  +i\frac{\sigma^{\mu\nu}k_{1\nu}\gamma^5}{2m}\widetilde{F}_8 \right). \label{eq:MGamma}
	\end{align}
	Here, the last diagram in the first equation stands for the counter term and $p_1 \cdot \epsilon_1 = 0$ is adopted. 
	The $ \widetilde{F}_i \, (i=1 \cdots 8)$ are functions of $t$ and masses, whose explicit definitions are presented at the end of this appendix.
	In Eq. \ref{eq:MGamma}, there are new form factors or new Lorentz structures other than $\gamma^\mu$ and $\sigma^{\mu\nu } p_{1\nu}$.
	The reason why we have such new Lorentz structures is that one of the tau lepton in above one-loop weak corrections to $\gamma\tau\tau$ vertex is not the on-shell spinor but an inner propagator, thus the Gordon identity fails there. 
	
	In conclusion, since the $\Gamma^\mu$ of the $\gamma\tau\tau$ vertex in $\gamma\gamma \longrightarrow \tau^+ \tau^-$ reaction contain new Lorentz structures, the feasibility to extract $a_\tau$ under simple $F_{1,2}$ parametrization scheme becomes in doubt. In addition, the corrections from irrelevant Feynman diagrams at higher order in such reaction also need to be suppressed.
	These can be checked by carrying out measurements on electron's or muon's AMM through the $\gamma\gamma \longrightarrow \l^+ \l^-$ reaction in heavy-ion or proton-proton UPC and making comparison with the precise measurements by the spin precession method. Note that several experiments have already observed the production of electron pairs \cite{PhysRevC.70.031902,PHENIX:2009xtn,ALICE:2013wjo,STAR:2019wlg} and muon pairs \cite{ATLAS:2020epq,CMS:2020skx} via photon-induced heavy-ion UPC, but do not try to extract their AMM.
	
	The explicit expressions of $ \widetilde{F}_i \, (i=1 \cdots 8)$ are as follows,
	\begin{equation}\label{eq:F1}
	\begin{aligned}
	\widetilde{F}_1 =&\frac{1}{2} \ln{\left[\frac{m^2}{m_W^2}\right]}+\frac{\left(\ln^2{\left[\frac{m_W^2}{m^2}\right]}-\ln^2{\left[-\frac{m_W^2}{t}\right]}+2
		\mathrm{Li_2} \left[\frac{m_W^2}{m^2}\right]-2 \mathrm{Li_2} \left[\frac{m_W^2}{t}\right]\right) \left(4 m^2-2 t+3 m_W^2\right)}{8 \left(m^2-t\right)}\\&+\frac{\ln{\left[\frac{m_W^2}{-m^2+m_W^2}\right]}
		\left(m^2-m_W^2\right) \left(m^4 \left(2 m^2+t\right)+m^2 \left(8 m^2+t\right) m_W^2+\left(-8 m^2+6 t\right) m_W^4\right)}{8 m^4 \left(m^2-t\right)
		m_W^2}\\&+\frac{3 m^4 \left(m^2-t\right) t+\left(2 m^6+m^4 \left(5-4 \pi ^2\right) t+m^2 \left(-7+2 \pi ^2\right) t^2\right) m_W^2+\left(4 m^4-m^2 \left(10+3
		\pi ^2\right) t+6 t^2\right) m_W^4}{8 m^2 \left(m^2-t\right) t m_W^2}\\&-\frac{\ln{\left[\frac{m_W^2}{-t+m_W^2}\right]} \left(m^2 t^2 \left(2 m^2+t\right)+2
		m^2 t \left(-2 m^2+5 t\right) m_W^2+\left(2 m^4-15 m^2 t+2 t^2\right) m_W^4+\left(4 m^2-2 t\right) m_W^6\right)}{8 \left(m^2-t\right) t^2 m_W^2},
	\end{aligned}
	\end{equation}
	
	\begin{equation}\label{eq:F2}
	\begin{aligned}
	\widetilde{F}_2 =&-\frac{1}{2} \ln{\left[\frac{m^2}{m_W^2}\right]}+\frac{\left(\ln^2{\left[\frac{m_W^2}{m^2}\right]}-\ln^2{\left[-\frac{m_W^2}{t}\right]}+2
		\mathrm{Li_2} \left[\frac{m_W^2}{m^2}\right]-2 \mathrm{Li_2} \left[\frac{m_W^2}{t}\right]\right) \left(2 t-3 m_W^2\right)}{8 \left(m^2-t\right)}\\&-\frac{\ln{\left[\frac{m_W^2}{-t+m_W^2}\right]}
		\left(t-m_W^2\right) \left(m^2 t \left(2 m^2+t\right)-\left(2 m^4+5 m^2 t\right) m_W^2+\left(4 m^2-2 t\right) m_W^4\right)}{8 \left(m^2-t\right)
		t^2 m_W^2}\\&+\frac{\ln{\left[\frac{m_W^2}{-m^2+m_W^2}\right]} \left(m^2-m_W^2\right) \left(m^4 \left(2 m^2+t\right)+\left(-8 m^4+m^2 t\right)
		m_W^2+2 t m_W^4\right)}{8 m^4 \left(m^2-t\right) m_W^2}\\&+\frac{m^4 \left(m^2-t\right) t+\left(2 m^6-7 m^4 t+m^2 \left(5-2 \pi ^2\right) t^2\right)
		m_W^2+\left(-4 m^4+m^2 \left(2+3 \pi ^2\right) t+2 t^2\right) m_W^4}{8 m^2 \left(m^2-t\right) t m_W^2},
	\end{aligned}
	\end{equation}
	
	\begin{equation}\label{eq:F3}
	\begin{aligned}
	\widetilde{F}_3 =&\frac{3 m \ln^2{\left[\frac{m_W^2}{m^2}\right]}}{8 \left(m^2-t\right)}-\frac{3 m \ln^2{\left[\frac{m_W^2}{t}\right]}}{8 \left(m^2-t\right)}+\frac{3
		m \mathrm{Li_2} \left[\frac{m_W^2}{m^2}\right]}{4 \left(m^2-t\right)}-\frac{3 m \mathrm{Li_2} \left[\frac{m_W^2}{t}\right]}{4 \left(m^2-t\right)}\\&+\frac{\ln{\left[\frac{m_W^2}{-m^2+m_W^2}\right]}
		\left(3 m^2-3 m_W^2\right)}{4 m^3-4 m t}+\frac{3 m \ln{\left[\frac{m_W^2}{m_W^2-t}\right]} \left(-t+m_W^2\right)}{4 \left(m^2-t\right) t},
	\end{aligned}
	\end{equation}
	
	\begin{equation}\label{eq:F4}
	\begin{aligned}
	\widetilde{F}_4 =&-\frac{m \left(-5 m^2+t\right) \ln{\left[\frac{m_W^2}{-t+m_W^2}\right]} \left(t-m_W^2\right)}{4 \left(m^2-t\right)^2 t}\\&+\frac{m \left(m^2
		\left(-8+5 \pi ^2\right)+\left(8-3 \pi ^2\right) t+2 \pi ^2 m_W^2\right)}{8 \left(m^2-t\right)^2}\\&-\frac{m \left(
		\ln^2{\left[\frac{m_W^2}{m^2}\right]} \left(5 m^2-3 t+2 m_W^2\right)-\ln^2{\left[-\frac{m_W^2}{t}\right]}
		\left(5 m^2-3 t+2 m_W^2\right)\right)}{8 \left(m^2-t\right)^2}
	\\&+\frac{m \left(-\mathrm{Li_2} \left[\frac{m_W^2}{m^2}\right] \left(5 m^2-3 t+2 m_W^2\right)+\mathrm{Li_2} \left[\frac{m_W^2}{t}\right]
		\left(5 m^2-3 t+2 m_W^2\right)\right)}{4
		\left(m^2-t\right)^2}\\&-\frac{m \left(\left(7 m^2-3 t\right) \ln{\left[\frac{m_W^2}{-m^2+m_W^2}\right]} \left(1-\frac{m_W^2}{m^2}\right)\right)}{4
		\left(m^2-t\right)^2},
	\end{aligned}
	\end{equation}
	
	\begin{equation}\label{eq:F5}
	\begin{aligned}
	\widetilde{F}_5 =&-\frac{\left(\ln^2{\left[\frac{m_W^2}{m^2}\right]}-\ln^2{\left[-\frac{m_W^2}{t}\right]}+2 \mathrm{Li_2} \left[\frac{m_W^2}{m^2}\right]-2
		\mathrm{Li_2} \left[\frac{m_W^2}{t}\right]\right) \left(3 m^2-t+3 m_W^2\right)}{4 \left(m^2-t\right)^2}\\&-\frac{\ln{\left[\frac{m_W^2}{-m^2+m_W^2}\right]}
		\left(1-\frac{m_W^2}{m^2}\right) \left(m^4+\left(9 m^2-2 t\right) m_W^2+2 m_W^4\right)}{4 \left(m^2-t\right)^2 m_W^2}\\&-\frac{\ln{\left[\frac{m_W^2}{-t+m_W^2}\right]}
		\left(t-m_W^2\right) \left(-m^4 t+\left(m^4-8 m^2 t\right) m_W^2+2 \left(m^2-2 t\right) m_W^4\right)}{4 \left(m^2-t\right)^2 t^2 m_W^2}\\&+\frac{m^2
		t \left(-m^2+t\right)-\left(m^4+m^2 \left(5-3 \pi ^2\right) t+\left(-6+\pi ^2\right) t^2\right) m_W^2+\left(-2 m^2+\left(2+3 \pi ^2\right) t\right)
		m_W^4}{4 \left(m^2-t\right)^2 t m_W^2},
	\end{aligned}
	\end{equation}
	
	\begin{equation}\label{eq:F6}
	\begin{aligned}
	\widetilde{F}_6 =&\frac{\left(\ln^2{\left[\frac{m_W^2}{m^2}\right]}-\ln^2{\left[-\frac{m_W^2}{t}\right]}+2 \mathrm{Li_2} \left[\frac{m_W^2}{m^2}\right]-2
		\mathrm{Li_2} \left[\frac{m_W^2}{t}\right]\right) \left(m^2-t+3 m_W^2\right)}{4 \left(m^2-t\right)^2}\\&-\frac{\ln{\left[\frac{m_W^2}{-m^2+m_W^2}\right]}
		\left(1-\frac{m_W^2}{m^2}\right) \left(m^4+\left(-7 m^2+2 t\right) m_W^2-2 m_W^4\right)}{4 \left(m^2-t\right)^2 m_W^2}\\&+\frac{\ln{\left[\frac{m_W^2}{-t+m_W^2}\right]}
		\left(t-m_W^2\right) \left(m^4 t-\left(m^4+4 m^2 t\right) m_W^2+2 \left(m^2-2 t\right) m_W^4\right)}{4 \left(m^2-t\right)^2 t^2 m_W^2}\\&+\frac{m^2
		t \left(-m^2+t\right)-\left(m^2-t\right) \left(m^2+\left(-6+\pi ^2\right) t\right) m_W^2+\left(2 m^2-\left(2+3 \pi ^2\right) t\right) m_W^4}{4 \left(m^2-t\right)^2
		t m_W^2},
	\end{aligned}
	\end{equation}
	
	\begin{equation}\label{eq:F7}
	\begin{aligned}
	\widetilde{F}_7 =&\frac{m^2 \left(-\ln^2{\left[\frac{m_W^2}{m^2}\right]}+\ln^2{\left[-\frac{m_W^2}{t}\right]}-2 \mathrm{Li_2} \left[\frac{m_W^2}{m^2}\right]+2
		\mathrm{Li_2} \left[\frac{m_W^2}{t}\right]\right)}{2 \left(m^2-t\right)}\\&+\frac{\frac{m^2 \left(-m^2+t+\pi ^2 t\right)}{m^2-t}-2 m_W^2}{2 t}-\frac{\ln{\left[\frac{m_W^2}{-m^2+m_W^2}\right]}
		\left(1-\frac{m_W^2}{m^2}\right) \left(m^4+3 m^2 m_W^2-2 m_W^4\right)}{2 \left(m^2-t\right) m_W^2}\\&+\frac{m^2 \ln{\left[\frac{m_W^2}{-t+m_W^2}\right]}
		\left(t-m_W^2\right) \left(m^2 t-m_W^2 \left(m^2-4 t+2 m_W^2\right)\right)}{2 \left(m^2-t\right) t^2 m_W^2},
	\end{aligned}
	\end{equation}
	
	\begin{equation}\label{eq:F8}
	\begin{aligned}
	\widetilde{F}_8 =&\frac{m^2 \left(-\ln^2{\left[\frac{m_W^2}{m^2}\right]}+\ln^2{\left[-\frac{m_W^2}{t}\right]}-2 \mathrm{Li_2} \left[\frac{m_W^2}{m^2}\right]+2
		\mathrm{Li_2} \left[\frac{m_W^2}{t}\right]\right)}{2 \left(m^2-t\right)}\\&+\frac{\frac{m^2 \left(-m^2+t+\pi ^2 t\right)}{m^2-t}+2 m_W^2}{2 t}-\frac{\ln{\left[\frac{m_W^2}{-m^2+m_W^2}\right]}
		\left(1-\frac{m_W^2}{m^2}\right) \left(m^4-m^2 m_W^2+2 m_W^4\right)}{2 \left(m^2-t\right) m_W^2}\\&+\frac{\ln{\left[\frac{m_W^2}{-t+m_W^2}\right]}
		\left(t-m_W^2\right) \left(m^4 t-m^4 m_W^2+2 m^2 m_W^4\right)}{2 \left(m^2-t\right) t^2 m_W^2}.
	\end{aligned}
	\end{equation}
	
\end{appendix}

\bibliography{ref}

\end{document}